\documentclass[12pt]{article}
\usepackage{amscd,amssymb,amsmath,latexsym,enumerate}
\usepackage[mathscr]{euscript}
\usepackage{epsfig}
\usepackage{fancybox}
\usepackage{verbatim}

\usepackage{tikz}
\usepackage{tikz-cd}
\usepackage{todonotes}

\usepackage{color}

\usepackage{xcolor}
\definecolor{MyBlue}{cmyk}{1,0.13,0,0.63}
\definecolor{MyGreen}{cmyk}{0.91,0,0.88,0.52}
\newcommand{\mylinkcolor}{MyBlue}
\newcommand{\mycitecolor}{MyGreen}
\newcommand{\myurlcolor}{black}

\usepackage{hyperref}
\hypersetup{%
  bookmarksnumbered=true,bookmarksopen=false,%
  plainpages=false,% necessary to prevent duplicate page identifiers
  linktocpage=true,%
  colorlinks=true,breaklinks=true,%
  linkcolor=\mylinkcolor,citecolor=\mycitecolor,urlcolor=\myurlcolor,%
  pdfpagelayout=OneColumn,%
  pageanchor=true,%
}

\title{A spectral localizer approach to strong topological invariants in the mobility gap regime}

\author{Tom Stoiber
\\
\\
{\small Yeshiva University, New York, USA}
}

\vspace{.75cm}

\date{\today}

\newtheorem{theorem}{Theorem}[section]

\newtheorem{definition}{Definition}[section]
\newtheorem{proposition}{Proposition}[section]
\newtheorem{lemma}{Lemma}[section]
\newtheorem{corollary}{Corollary}[section]

\newcommand{\CM}{{\mathbb C}}
\newcommand{\NM}{{\mathbb N}}
\newcommand{\RM}{{\mathbb R}}

\newcommand{\TM}{{\mathbb T}}
\newcommand{\ZM}{{\mathbb Z}}
\newcommand{\PM}{{\mathbb P}}

\newcommand{\PP}{{\bf P}}

\newcommand{\DD}{{\bf D}}

\newcommand{\Aa}{{\cal A}}
\newcommand{\Bb}{{\cal B}}

\newcommand{\Ee}{{\cal E}}

\newcommand{\Ss}{{\cal S}}

\newcommand{\Cc}{{\cal C}}

\newcommand{\Ll}{{\cal L}}

\newcommand{\Hh}{{\cal H}}

\def\esssup{\mathop{\rm ess\,sup}}

\newcommand{\one}{{\bf 1}}

\newcommand{\Ind}{{\rm Ind}} 
\newcommand{\Ker}{{\rm Ker}}

\newcommand{\sgn}{{\rm sgn}} 
\newcommand{\Sig}{{\rm Sig}}

\newcommand{\bbC}{\mathbb{C}}

\newcommand{\bbN}{\mathbb{N}}
\newcommand{\bbP}{\mathbb{P}}

\newcommand{\bbR}{\mathbb{R}}

\newcommand{\bbZ}{\mathbb{Z}}

\newcommand{\symlabel}{\sigma}

\newcommand{\difd}{\textup{d}}

\providecommand{\abs}[1]{\left \lvert#1 \right \rvert} 
\providecommand{\norm}[1]{\left \lVert#1 \right \rVert}
\DeclareMathOperator*{\slim}{s-lim}

\newcommand{\Dom}{\mathrm{Dom}}

\begin{document}

\maketitle

%%%%%%%%%%%%%%%%%%%%%%%%%%%%%%%%%%%%%%%%%%%%%%%%%%%%
\begin{abstract}
Topological phases of gapped one-particle Hamiltonians with (anti)-unitary symmetries are classified by strong topological invariants according to the Altland-Zirnbauer table. Those indices are still well-defined in the regime of strong disorder when the spectral gap is replaced by a mobility gap, however, many questions regarding their robustness and existence of topological boundary states are wide open. 
We apply the recently developed spectral localizer method to prove results on the stability of strong topological invariants under a notion of continuous homotopy that preserves a mobility gap condition. Using the local computability afforded by the spectral localizer we show that for parametrized random families that satisfy a fractional moments bound the probability distribution of the strong topological invariant changes continuously. In particular, for ergodic families the almost sure index must be constant on any path which preserves the mobility gap. Using similar methods, we also prove a result on the delocalization of interface states between two mobility-gapped systems which have differing strong invariants.
\end{abstract}

%%%%%%%%%%%%%%%%%%

%%%%%%%%%%%%%%%%%%%%%%%%%%%%%%%%%%
\section{Introduction}

It is well-accepted that gapped topological phases of free fermions with short-range hopping are classified up to stable homotopy by topological indices depending on their Altland-Zirnbauer(AZ) symmetry class \cite{AZ}. As highlighted by Kitaev \cite{Kitaev} this can be understood via K-theory, in particular for translation-invariant models the eigenspaces of gapped Hamiltonians define vector bundles over some momentum space, which are up to stable homotopy classified by topological K-theory. The use of the non-commutative version of the theory, K-theory for $C^*$-algebras, in solid-state physics predates those developments \cite{Bel}, and famously lead to an explanation of the integer Quantum Hall effect \cite{BES} and the edge states thereof \cite{KellendonkRMP2002}, which is the special case of the complex symmetry class $A$ in the Altland-Zirnbauer classification. More recently, the K-theoretic and operator-algebraic approach to the classification of topological insulators and their bulk-boundary correspondence has been extended to cover disordered matter in all Altland-Zirnbauer classes and with possible additional crystalline symmetries (see in particular \cite{FreedAHP2013, ThiangAHP16,
PSbook,Prodan2017,Kellendonk17,Kubota, EwertMeyer,BKR,AMZ,BM19}). 

Let us briefly recall how the operator-algebraic approach connects to the AZ-classification (for details we refer to \cite{Kellendonk17,Kubota,EwertMeyer}). One assumes that a gapped Hamiltonian $H$ on a discrete pattern $\Ll\subset \RM^d$ acts on the Hilbert space $\ell^2(\Ll)\otimes \CM^n$, $n$ the number of on-site degrees of freedom, and (for simplicity) has finite hopping range, i.e. $\langle x| H|y\rangle = 0$ whenever $\abs{x-y}>R$ for some $R$. One can classify such a gapped Hamiltonian $H$ by its band-flattening $\sgn(H)$ which is a self-adjoint unitary and defines an element of the Roe $C^*$-algebra $C^*_{\mathrm{Roe}}(\Ll)$, a universal $C^*$-algebra generated by finite-hopping-range operators on $\Ll$. According to the selected AZ-class the Hamiltonian $H$ and thus $\sgn(H)$ must satisfy specified (anti-)commutation relations with respect to certain symmetry operators, which represent time-reversal, particle-hole or chiral symmetry. Due to the relations of those symmetries with real Clifford algebras, the stable homotopy classes of such self-adjoint unitaries are precisely classified by the complex K-groups $K_j(C^*_{\mathrm{Roe}}(\Ll))$ if the symmetries are unitary and in the anti-unitary case by the real K-groups $KR_j(C^*_{\mathrm{Roe}}(\Ll))$ with $j\in \ZM_2$ or $j\in \ZM_8$ respectively. Those options for $j$ are in one-to-one correspondence to the symmetry types in the Altland-Zirnbauer classification. If $\Ll$ is coarsely equivalent to $\RM^d$, e.g. if it is a Delone set, then
$$K_j(C^*_{\mathrm{Roe}}(\Ll))=\begin{cases} \ZM & \text{if }j=d\;\mathrm{mod}\; 2\\
0 & \text{else}\end{cases}, \quad KR_j(C^*_{\mathrm{Roe}}(\Ll))=
\begin{cases} 
\ZM & \text{if }j-d \mod 8 \in \{0,4\}\\
\ZM_2 & \text{if }j-d\mod 8 \in \{1,2\}\\
0 & \text{else.}\end{cases}$$
The index in $\ZM$ or $\ZM_2$ associated to the Hamiltonian is called its {\it strong topological invariant}. It arises as the pairing of the class defined by a Hamiltonian with a K-homology class defined by the geometry, namely the class of a position-space Dirac operator $D_\Ll$ of the pattern. If $j$ labels the symmetry type of the Hamiltonian and $d$ the dimension of space then such a pairing takes values in $K_{j-d}(\CM)$ for unitary symmetries, which is either $0$ or $\ZM$, whereas for anti-unitary symmetries the possible values $KR_{j-d}(\CM)$ are $0$, $\ZM$ or $\ZM_2$. Two gapped Hamiltonians can be (after stabilization with additional on-site degrees of freedom) connected by a symmetry-preserving homotopy consisting of gapped Hamiltonians with short hopping range  if and only if they have the same $\ZM$- respectively $\ZM_2$-index. This provides a mathematical justification for the Altland-Zirnbauer classification.

The Roe $C^*$-algebras are very large and only take short hopping range into account, moreover, some of the homotopies may be difficult to realize in a physical system. Depending on the physical situation one attempts to model it is therefore often preferable to use more specialized $C^*$-algebras for the Hamiltonians such as groupoid- or crossed product algebras  \cite{PSbook,BP, BM19}. In that case the K-groups and thus the homotopy-classification can also be much richer, for example, include additional so-called weak topological invariants that are not captured by the AZ-classification. With respect to such a more restricted notion of homotopy the strong topological invariant still provides an obstruction to the existence of gapped homotopies but is not a complete invariant anymore.

Those strong topological invariants are robust under any localized perturbation as well as extended perturbations which do not close the spectral gap. For their definition one does not actually require the existence of a spectral gap, instead it is sufficient for the Hamiltonian to have a mobility gap (see e.g. \cite{GS16}), which means that the system is an insulator not because there is an energy gap at the Fermi level but instead because the spectral region around the Fermi level is made up entirely of (Anderson-)localized states. This will generally be the case in the regime of moderate to strong random disorder, which is strong enough to close all spectral gaps, but not necessarily strong enough to completely localize the entire spectrum. It is difficult to apply $K$-theory directly in this case. For one, there are no similarly convenient algebras which are known to contain the band-flattenings of mobility-gapped Hamiltonians like the Roe $C^*$-algebras (those do have short hopping range in some sense, but not in the uniform sense that would be required). Moreover, the map from the Hamiltonian to its band-flattening is not continuous in operator-norm if there is no spectral gap, which is problematic since K-theory is based on norm-continuous homotopies. 

Instead of K-theory, existing approaches are therefore more focused on directly defining the strong topological invariant, e.g. as a Fredholm index, and studying its properties (e.g. \cite{BES,RSb,GS16, PSbook, SSt, Shapiro, CS23}). Not much is known about the homotopy classification in the mobility-gapped case, in particular, it is not even known in general if homotopies that preserve the mobility gap also preserve the strong topological invariant. There do exist partial results: In more than half of the Altland-Zirnbauer classes the index is discrete and can be computed by a local formula as a so-called real-space Chern number. Thinking of the index as being associated to random ergodic families of Hamiltonians, the Chern numbers are almost surely constant on paths which preserve the mobility gap \cite{PSbook}. This is a combination of two facts: The index is integer-valued and is computed by a local expression that manifestly depends continuously on the Fermi projections in a sufficiently strong topology, namely that of a non-commutative Sobolev space. A similar result can be shown under a strong deterministic localization condition \cite{Shapiro}. Those approaches do not generalize easily to those Altland-Zirnbauer classes which are classified by $\ZM_2$-valued indices since those are not computable by a similarly convenient local real-space formula.

However, there is another recent method to compute the index pairings  arising from a Dirac operator and self-adjoint gapped Hamiltonians: The spectral localizer method \cite{LS,LS2,LS3,LSS, ST2021, DS21} expresses those topological invariants in terms of spectral invariants of finite-dimensional invertible matrices which have similar homotopy invariance as the indices themselves. In particular, the method has recently been extended to $\ZM_2$-invariants, which are then computed as the signs of determinants or Pfaffians of finite-dimensional matrices \cite{DS21}. The main idea of this paper is to use those finite-volume expressions as local index formulas to study mobility-gapped phases of matter that are classified by $\ZM_2$-invariants.

The first main result uses those numerical expressions for the index pairings to prove stability for the strong topological invariant  under some notion of homotopy for mobility-gapped Hamiltonians. One challenge is that the standard assumptions of the spectral localizer method do not apply in the mobility-gapped regime, since the Hamiltonian is not an invertible operator. We circumvent this by spectral flattening, i.e. instead of a physical Hamiltonian $h$ we directly use its band-flattening $H=\sgn(h)$ as input for the spectral localizer. The band-flattenings of a parametrized family of mobility-gapped Hamiltonians, however, will usually not depend continuously on the parameters in any relevant operator topology. For this reason we will instead of single Hamiltonians consider ensembles of random operators, such that the band-flattenings can satisfy some sort of averaged continuity in the sense that large jumps of their matrix elements are rare events. In this work we follow \cite{PSbook,Prodan2017} and axiomatize the mobility-gap regime by imposing a fractional moments bound \cite{AizenmanMolchanov}. We state it in a similar form as in \cite{Prodan2017} such that it subsumes a Combes-Thomas estimate for all energies outside the spectrum:
\begin{definition}
\label{def:mbg}
Let $(\Omega,\PM)$ be a probability space with $\Omega$ a compact Hausdorff space and $\PM$ a regular Borel measure. Let $h=(h_\omega)_{\omega\in \Omega}$ with $h_\omega \in \Bb(\ell^2(\Ll) \otimes \bbC^n)$ be a measurable family of bounded self-adjoint operators for $\Ll\subset \RM^d$ a uniformly discrete set.

We say that $h$  has a mobility gap in the interval $\Delta$ if there is some fractional exponent $0 < \nu < 1$ such that for each $k\in \bbN$ and any small enough $\delta>0$ there is a constant $A_k(\delta) > 0$ such that
\begin{equation}
\label{eq:fractional_moments}
\sup_{x,y\in \Ll}\langle x-y\rangle^{k} \int_{\Omega} \norm{\langle x|\frac{1}{h_\omega-\eta}|y\rangle}_{M_n(\CM)}^\nu \difd\mathbb{P}(\omega) <  A_k(\delta) \, 
\end{equation}
for all $\eta \in \CM\setminus \RM$ with $\mathrm{dist}(\eta, \mathrm{Spec}(h)\setminus \Delta) > \delta$, where one uses the operator-norm on $M_n(\bbC)$ and as usual $\langle x\rangle = (1+x^2)^{\frac{1}{2}}$.
\end{definition}
We then obtain the following continuity result for the probability distributions of the strong topological invariant:
\begin{theorem}
\label{th:main}
Let $T$ be a metric space, $\Ll\subset \RM^d$ a Delone set and $(\Omega,\PM)$ a probability space as in Definition~\ref{def:mbg}. We assume that $t \in T \mapsto h(t)$ is a parametrized family of random Hamiltonians on $\ell^2(\Ll)\otimes \CM^n$, i.e. for each $t$ one has a measurable family $h(t)=(h_\omega(t))_{\omega \in \Omega}$ of self-adjoint operators, with the following properties: \begin{enumerate}
\item[(i)] The matrix elements $\langle x| h_\omega(t)|y\rangle \in M_n(\CM)$ decay faster than any polynomial
$$\sup_{t\in T, \omega\in\Omega, x,y\in \Ll} \langle x-y\rangle^k \norm{\langle x| h_\omega(t)|y\rangle}_{M_n(\CM)} <\infty, \qquad \forall k\in \NM.$$
\item[(ii)] The matrix elements as random variables are continuous in the averaged sense that
$t \in T\mapsto \langle x| h(t)| y\rangle \in L^2(\Omega,M_n(\CM))$ is continuous for all $x,y\in \Ll$.
\item[(iii)] There is an interval $\Delta \subset \RM$ with $0\in \Delta$ which is a mobility gap for each random family $h(t)$ with parameters that can be chosen uniformly in $t$.
\item[(iv)] For each $t\in T$ one has that $0$ is almost surely not an eigenvalue of $h_\omega(t)$.
 \item[(v)] For each $t\in T$ and almost every $\omega$ the Hamiltonian $h_\omega(t)$ satisfies the symmetry relations appropriate for some Altland-Zirnbauer class $\symlabel$ and hence one can define the strong topological invariant as
 $$\Ind_\symlabel(F_\omega(t)):= \Ind(F_\omega(t))\in \ZM$$
 if it is supposed to be integer-valued in the respective symmetry class and otherwise
 $$\Ind_\symlabel(F_\omega(t)):=  \Ind_2(F_\omega(t))\in \ZM_2.$$
 Here $F_\omega(t)$ is a Fredholm operator constructed from $\DD$ and the band-flattening $\sgn(h_\omega(t))$ appropriate for the symmetry class $\sigma$.
\end{enumerate} 
Then the probability $$t\in T \mapsto \PM(\Ind_\symlabel(F_\omega(t))=z)$$ that the index takes any particular value $z\in \ZM$ respectively $z\in \ZM_2$ is a continuous function.
\end{theorem}
The unexplained notions will be made precise in the main text. A simple sufficient condition for (ii) which is enough for most cases of interest is continuity of each map $t\in T\mapsto \langle x|h_\omega(t)|y\rangle$ with a modulus of continuity that is bounded uniformly in terms of $\omega$. In the situation of a mobility gap one expects dense point spectrum, hence (iv) should always hold if $0$ was just a generic spectral value. However, in some of the symmetry classes the spectrum of the Hamiltonian is symmetric around $0$ and that special value can turn out to be an eigenvalue with positive probability, which we need to exclude by assumption since it makes the topological invariant ill-defined. We are mostly interested in the case $T=[0,1]$ which turns this into a statement about homotopies of Hamiltonians that preserve the mobility gap. Note that we can only prove that the probability distributions are continuous, not that the indices themselves are almost surely constant under homotopy. There are, however, important situations in which the index for any fixed $t$ is known to be the same for almost every realization $\omega$, in particular when the Hamiltonians are ergodic under translations. In that case, this almost sure index will be actually constant on any path which preserves the mobility-gap condition, see Section~\ref{sec:tr_inv}.

The subject of the final Section~\ref{sec:interfaces} will be the second main result, also proved using the local computability provided by the skew localizer, which we roughly state as follows:
\begin{theorem}
Let $h^{(I)}=(h^{(I)}_\omega)_{\omega\in \Omega}$ be a random family of Hamiltonians representing an interface between two mobility-gapped Hamiltonians $h^{(1)}=(h^{(1)}_\omega)_{\omega\in \Omega}$ and $h^{(2)}=(h^{(2)}_\omega)_{\omega\in \Omega}$. If the three Hamiltonians are of the 
same Altland-Zirnbauer symmetry type $\symlabel$ and almost surely none of them has $0$ as an  eigenvalue then
$$\PM(\Ind_\symlabel(F_\omega^{(1)}) \neq \Ind_\symlabel(F_\omega^{(2)}))>0$$
implies that $h^{(I)}$ does not have a mobility gap in any interval around $0$.
\end{theorem}
More precise conditions will be given in Section~\ref{sec:interfaces}.  This result provides evidence that interfaces of mobility-gapped strong topological insulators must have conducting interface states and generalizes similar recent results that only apply to a subset of the Altland-Zirnbauer symmetry classses, namely those protected by integer-valued strong invariants \cite{PSbook,BolsWerner,SSt} as well as the time-reversal symmetric classes in two dimensions \cite{BolsCedzich}.

\section{Index pairings with a Dirac operator}

In this section we recall how to define the strong topological invariants for the Altland-Zirbauer classes in terms of indices of appropriate Fredholm operators. There are different but equivalent approaches and perspectives one can assume, especially when it comes to the real (anti-unitary) symmetry relations required to describe physical systems with time-reversal or particle-hole symmetries. For example, one can use different equivalent formalisms of real $K$-theory, in particular $KO$-theory \cite{BKR}, the van-Daele-picture of KR-theory \cite{Kellendonk17} or twisted equivariant K-theory \cite{Kubota}. In this paper we instead follow the operator-theoretic viewpoint of \cite{GS16} where the indices are defined as Fredholm indices of operators with certain real symmetries, since that is the formalism that the skew-localizer method \cite{DS21} is built upon. This is also well-adapted to the mobility-gapped situation we are interested in, since many of the purely K-theoretic arguments break down or at least lead to major technical problems when considering Hamiltonians without a spectral gap, while an index-based approach is still meaningful. Most of this section is just review based on material from \cite{GS16}.

A central object is a (dual) Dirac operator that encodes the geometry of real-space pattern in the sense of spectral geometry:
\begin{definition}
\label{def:standarddirac}
A $d$-dimensional Dirac operator is an operator of the form
$$\DD  = \sum_{i=1}^d \gamma_i \otimes D_i$$
where $D_1,...,D_i$ are real commuting self-adjoint operators acting on a common core contained in a Hilbert space $\Hh_0$ and $\gamma_1,...,\gamma_d \in M_{d'}(\bbC)$, $d'=2^{\lfloor d/2\rfloor}$ are the representation of the complex Clifford algebra with $d$ generators defined in Appendix~\ref{sec:app}. It is further assumed that $\sum_{i=1}^n D_i^2$ is invertible and has compact resolvents, thus the same is true for $\DD$.
\end{definition}
The complex Hilbert space $\Hh_0$ shall have a real structure and the representation space of the Clifford algebra also has a fixed complex conjugation. In the applications $\Hh_0$ will be $\ell^2(\Ll)$ for a uniformly discrete set $\Ll\subset\RM^d$ and $\DD$ is built from the position operators of the pattern, see Section~\ref{sec:mbg}. If $d$ is even then $\DD$ anti-commutes with the symmetry operator $$\gamma_0 = \imath^{d/2} \gamma_1\, \dots \gamma_d$$ and one can understand the anti-commutation relation as $\DD$ being odd in the grading provided by $\gamma_0$, i.e. when decomposing $\CM^{d'}$ into $P_\pm = \frac{1}{2}(\one \pm \gamma_0) \CM^{d'}$ then $\DD$ becomes an off-diagonal matrix
$$\DD = \begin{pmatrix}
	0 & D_0^* \\ D_0 & 0
\end{pmatrix}, \qquad \sgn(\DD) = \begin{pmatrix}
0 & G^* \\ G & 0
\end{pmatrix}.$$
For the representation as defined in Appendix~\ref{sec:app}, $\gamma_0$ is diagonal with $\gamma_0=\one_{{d'}/2}\oplus (-\one_{{d'}/2})$. If $\DD$ is invertible, as we will also assume throughout, then the off-diagonal parts $G,G^*$ are therefore unitary operators on $\Hh_0\otimes \CM^{{d'}/2}$. 

The Dirac operator encodes the real-space geometry into the Hilbert space and the compactness of the resolvent expresses that this real-space geometry corresponds to a uniformly discrete pattern. The time evolution of a physical system in the one-particle picture is described by a self-adjoint Hamiltonian which is matrix-valued with an $n$-dimensional space of on-site degrees of freedom. Hamiltonians $H$ and Dirac operators $\DD$ act on the common Hilbert space $\Hh=\Hh_0\otimes \CM^{d'}\otimes \CM^n$ by amplifying their matrix degrees of freedom with suitable trivial tensor products, which will be suppressed in the notation.
\begin{definition}
\label{def:hamiltonian}
A Hamiltonian $H \in \Bb(\Hh_0\otimes \CM^n)$ shall be a self-adjoint operator for which $0$ is not an eigenvalue of $H$ and the commutator $[\sgn(H), \sgn(\DD)]$ is a compact operator.

We say that $H$ is more strongly a gapped short-range Hamiltonian if
\begin{enumerate}
\item[(i)] $H$ is a self-adjoint invertible operator.
\item[(ii)] $H$ maps the domain of $\DD$ to itself and $[\DD, H]$ extends to a bounded operator.
\end{enumerate}
\end{definition}
The compactness of the commutator $[\sgn(H), \sgn(\DD)]$ is the fundamentally important requirement that makes the topological invariants well-defined. It is natural from the perspective of non-commutative geometry \cite{Connes}, but also alternatively argued for in \cite{Shapiro}. More generally, compactness of the commutator $[T, \sgn(\DD)]$ between a self-adjoint operator $T$ and the sign of the Dirac operator (sometimes called the quantum differential) and boundedness of $[\DD,T]$ both are abstract conditions on the decay rate of the matrix elements of $T$ with respect to the geometry encoded into $\DD$.

The band-flattened Hamiltonian $\sgn(H)$ or equivalently the Fermi projection $\chi(H<0)$ onto the spectral region below the Fermi energy, which we by convention fix at $0$, encodes the ground state of the free-fermion system. It is well-known \cite[Lemma 10.1.8]{GVF} that (i) and (ii) together with compactness of the resolvent of $\DD$ imply the compactness of $[H,\sgn(\DD)]$ by functional calculus of $\DD$ which implies in turn compactness of $[\sgn(H),\sgn(\DD)]$. However, the commutator $[\sgn(H),\sgn(\DD)]$ can be compact under much weaker conditions. In particular, this will be true even if $H$ does not have a spectral gap, but only a so-called mobility gap. Conversely, the condition does presumably never hold for metallic systems with conducting states at the Fermi energy, therefore the systems under consideration in this work are all (topological) insulators.

Half of the Altland-Zirnbauer classes involve a chiral symmetry of the Hamiltonian:
\begin{definition}
\label{def:chiralsymmetry}
A Hamiltonian $H \in \Bb(\Hh_0 \otimes \CM^n)$ has a chiral symmetry if there is a self-adjoint unitary $J \in M_n(\CM)$ which anti-commutes with $H$, i.e. $J H J = -H$, and is balanced in the sense that $J\simeq -J$ (unitary equivalence).

In that case $\Hh_0\otimes \CM^n$ decomposes into the two eigenspaces $\Hh_+ \oplus \Hh_-$ of $J$ for eigenvalues $\pm 1$ and written as a matrix one has
$$H = \begin{pmatrix}
	0 & H_0^* \\ H_0 & 0
\end{pmatrix}.$$
We will say that $J$ is the standard chiral symmetry if $n$ is even and $J=\one_{n/2} \oplus (-\one_{n/2})$, in which case $H_0$ and $H_0^*$ both are canonically identified with operators mapping $\Hh_+\simeq \Hh_0\otimes \CM^{n/2}$ into itself.
\end{definition}
Any chiral symmetry can be made standard by a change of basis, but not canonically.

For an even-dimensional Dirac operator we extract the topological invariants from the unitary $G=D_0\abs{D_0}^{-1}$ by pairing it with the Fermi projection $P=\chi(H<0)$, while for an odd-dimensional Dirac operator we pair the Fermi unitary $U=H_0 \abs{H_0}^{-1}$ with the spectral projection $E=\chi(\DD > 0)$. The integer-valued index pairings are computed as the index $\Ind(T)=\dim \ker(T)-\dim \ker(T^*)$ of one of the Fredholm operators
\begin{equation}
\label{eq:freds1}
F_{pu}=E U E + \one - E, \qquad F_{up}=P G P + \one-P
\end{equation}
with the notation signifying a projection/unitary respectively a unitary/projection pairing. 
More precisely one needs to pad $U$, $P$ respectively $E$, $G$ by tensor products with some identity matrices to make them act on the same space, but we will suppress such trivial tensor products in the notation. One can also consider the projection/projection pairing
\begin{equation}
\label{eq:freds2}
F_{pp}=E (\one - 2 P) E + \one-E
\end{equation}
which defines a non-trivial Fredholm operator, but it always has index zero since it is self-adjoint. However, it may have a non-trivial secondary $\ZM_2$-valued index
$$\Ind_2(F_{pp}) = \mathrm{dim}\, \Ker(F_{pp})\mod 2$$
that is not generally homotopy-invariant without additional symmetries.

The $\ZM$-valued index pairings $\Ind(F_{up})$ in even dimension $d$ and $\Ind(F_{pu})$ in odd dimension if the Hamiltonian has a supplemental chiral symmetry are precisely the strong topological invariants of the complex (unitary) classes A respectively AIII in the Altland-Zirnbauer classification of topological insulators.

The real (anti-unitary) symmetry classes of the Altland-Zirnbauer table feature additional symmetries which relate Hamiltonians with their complex conjugates. We denote by $\Cc$ the complex conjugation on any Hilbert space with a real structure, i.e. an anti-linear involution  which defines a complex conjugation $\overline{S}=\Cc S \Cc^*$ for bounded operators $S$ and a transpose $S^t = \overline{S}^*$. 

On the space $\CM^n$ of on-site degrees of freedom one needs one or two commuting anti-unitary operators $\mathfrak{T}=T\Cc$ and  $\mathfrak{P}=P\Cc$, representing time-reversal and particle-hole conjugation respectively with unitary matrices $T$ and $P$ which may depend on the Hamiltonian. Depending on the symmetry type one wants to realize $\mathfrak{T}$, $\mathfrak{P}$ must be chosen to square to $\one$ or $-\one$, which enforces relations $T\overline{T}=\pm \one$ and $P\overline{P}=\pm \one$. We denote those two signs by $s_T,s_P \in \{-1,1\}$ respectively. 

\begin{definition}
	\label{def:symmetries}
	For $H=H^* \in \Bb(\Hh_0 \otimes \CM^n)$ one distinguishes real symmetry types according to the following table
	\begin{center}
		\begin{tabular}{c|c|c|c|c|c|c|c|c}
			$j$ & $0$ & $1$ & $2$ & $3$& $4$ &$5$ &$6$ &$7$ \\
			CAZ & AI & BDI & D & DIII & AII & CII & C & CI \\
			\hline
			$T^* \overline{H} T =  H$ & $1$ & $1$ & $0$ & $-1$ & $-1$ & $-1$ &  $0$ &  $1$\\
			\hline
			$P^* \overline{H} P = -H$ & $0$ & $1$ & $1$ &  $1$ &  $0$ & $-1$ & $-1$ & $-1$\\
		\end{tabular}
	\end{center}
	where $\pm 1$ means that the respective relation holds with $s_T=\pm 1$ respectively $s_P=\pm 1$, while a $0$ entry means that no such (anti-)symmetry is imposed upon $H$.
\end{definition}
We will use the Cartan-Altland-Zirnbauer (CAZ) label assigned to each symmetry-type equivalently to $j$, since it is easier to memorize, but its original meaning as the type of a symmetric space is not really relevant here. The index $j$ in turn is related to Real $K$-theory; the band-flattenings $\sgn(H)$ define classes in the Real $K$-group $\mathrm{KR}_j(\Aa)$ of a suitable $C^*$-algebra and there is a (not entirely canonical) isomorphism between the twisted equivariant K-groups which classify Hamiltonians with AZ-type symmetries and the Real K-theory groups $\mathrm{KR}_j$ \cite{Kubota}.

One can bring any $T$, $P$ into a standard form after possibly enlarging the dimension of the fiber $n$ (this can be shown similarly to \cite[Proposition 19]{GS16}). Enlarging the fiber dimension entails amplifying Hamiltonians with a direct sum of trivial Hamiltonians of the respective symmetry class, an operation which leaves K-theory classes and therefore indices invariant. This procedure is ultimately not important here, what is important is that fixing a standard choice for the symmetry operators does not restrict the possible topological phases of Hamiltonians in any way.
\begin{definition}
\label{def:standard_symmetries}
We say that the symmetry operators $T$, $P$ are in standard form for some $j\in \ZM_8$ if they are given by matrices in $M_n(\CM)$ which are tensor products of some identity matrix with \\
\begin{center}
 \begin{tabular}{c|c|c|c|c|c|c|c|c}
		$j$ & $0$ & $1$ & $2$ & $3$& $4$ &$5$ &$6$ &$7$ \\
		\hline
		$T$ & $\one$ & $\one$ &  & $\sigma_1\otimes \imath\sigma_2$ & $\imath \sigma_2$ & $\one \otimes \sigma_2$ &   &  $\sigma_1$\\
		\hline
		$P$ &   & $\sigma_3$ & $\one$ &  $\sigma_2\otimes \imath\sigma_2$ &   &  $\sigma_3 \otimes \sigma_2$ & $\imath \sigma_2$ & $\sigma_2$\\
	\end{tabular}\\
\end{center}
Denote by $\mathrm{Ham}_{j}(n)$ for $j \in \ZM_8$ and compatible fiber dimension $n$ the $\bbR$-linear space of self-adjoint bounded operators on $\Bb(\Hh\otimes \CM^n)$ which satisfy the symmetry relations of Definition~\ref{def:symmetries} with respect to the given symmetry operators. To unify the notation we also include the even/odd complex symmetry type by denoting furthermore $\mathrm{Ham}_{e}(n)$ the space of self-adjoint operators and $\mathrm{Ham}_{o}(n)$ the space of self-adjoint operators which anti-commute with the standard chiral symmetry, such that in total the labels can take values in $\symlabel\in \{e,o\} \cup \ZM_8$.
\end{definition}

With this choice the symmetry operators $\mathfrak{T}$ and $\mathfrak{P}$ commute and if $j$ is even then $TP$ is proportional to $\sigma_3 \otimes \one$, hence one of the symmetry relations can be written as $J H J=-H$ with the standard chiral symmetry $J$. This is required to unambiguously define the index pairings for the even symmetry classes, which require reducing out the off-diagonal part $H_0$ of $H$ as in Definition~\ref{def:chiralsymmetry}. The second symmetry relation can then be written in terms of $H_0$ and takes the form $S^* \overline{H_0} S= H_0$ if $s_Ps_T=1$ or $S^* H_0^t S= H_0$ if $s_Ps_T=-1$ for $S$ some real unitary. 

For the non-trivial cases $j-d\mod \{0,1,2,4\}$ one can define a $\ZM$-, $2\ZM$- or $\ZM_2$-valued index which classifies gapped Hamiltonians up to stable homotopy. For that it is important that the Dirac operator must also have compatible real symmetries, which is in our case ensured from the choice of Clifford representation that we choose as in \cite{GS16} respectively Appendix~\ref{sec:app}. Those symmetries imply that $\DD$ defines a real $KR$-homology class, respectively a real spectral triple of KO-dimension $d \mod 8$ \cite{GS16}. From the perspective of index pairings with $KR$-theory, the Dirac operator could be of a more general form than Definition~\ref{def:standarddirac} as long as it has the correct symmetry type. However, the standard form adopted here is well-adapted to satisfy all assumptions of the skew localizer method \cite{DS21}.

The integer-valued indices are still the indices of the above Fredholm operators \eqref{eq:freds1}, which may be symmetry-constrained to only take values in $2\bbZ$. In the $\ZM_2$-valued cases the symmetries constrain the Fredholm index of $F_{pu}$ respectively $F_{up}$ to vanish, but imply the stability of a supplementary $\bbZ_2$-valued index. In total one ends up with the Fredholm operators \eqref{eq:freds1} and \eqref{eq:freds2} and has possibly non-trivial index pairings depending on the symmetry-type of the Hamiltonian and dimension of the Dirac operator as follows:

\vspace{.25cm}
\begin{tabular}{c|c|c|c|c|c|c|c|c}
	$j-d \mod 8$ & $0$ & $1$ & $2$ & $3$ & $4$ & $5$ & $6$ & $7$ \\
	\hline
	$d$ even &  $\Ind(F_{up})$ &  $\Ind_2(F_{up})$ &  $\Ind_2(F_{up})$ & 0 & $\Ind(F_{up})$ & 0 & 0 & 0 \\
	$d$ odd & $\Ind(F_{pu})$ &  $\Ind_2(F_{pp})$ &  $\Ind_2(F_{pu})$ & 0 & $\Ind(F_{pu})$ & 0 & 0 & 0\\
\end{tabular}\\

\vspace{.25cm}
The entry in the column $4$ takes values in $2\ZM$. If one sets up the physically relevant Dirac operator and symmetry operators then this table captures precisely the Altland-Zirnbauer strong invariants of topological insulators.

\section{The spectral localizer}

The spectral localizer method, originally introduced in \cite{
LS} and subsequently extended and refined in \cite{LS2,LSS,ST2021,DS21,DLS} is a method to reduce the index pairing between a suitable Dirac operator and a gapped short-range Hamiltonian to the computation of spectral invariants of invertible finite-dimensional matrices. The fundamental idea is to combine $H$ and $\DD$ into a self-adjoint operator with compact resolvent, called the spectral localizer,
\begin{equation}
\label{eq:sl_big}
L_\kappa = \begin{pmatrix}
	\kappa \DD & H\\ H & -\kappa \DD
\end{pmatrix} = \kappa \sigma_3 \otimes \DD + \sigma_1 \otimes H
\end{equation}
with the Pauli matrices $\sigma_1,\sigma_3$. The underlying Hilbert space is $\CM^2\otimes \Hh = \CM^2\otimes (\Hh_0\otimes \CM^{d'} \otimes \CM^n)$ and we recall that we omit trivial tensor products. If $H$ has a spectral gap of size $g$ around $0$, which can be written as $H^2>g^2\one$, and $[\DD,H]$ is a bounded operator then the spectral localizer is invertible for small enough $\kappa$ as
$$L_\kappa^2 \geq  \kappa^2 \DD^2 + H^2 - \lVert \kappa [\DD,H]\rVert \one$$
has a strictly positive lower bound. Some more intricate estimation shows that the same is still true for the finite-volume restrictions
$$L_{\kappa, \rho} := \pi_\rho \circ L_\kappa \circ \pi_\rho^*$$
provided that $\kappa$ is first chosen small enough and then $\rho$ is chosen large enough. Here, $\pi_\rho$ denotes the restriction to the image of the spectral projection $\chi(\abs{\DD}\leq \rho)$ on any Hilbert space where that makes sense. Since the spectral gap is shared between $H$ and $L_{\kappa,\rho}$ for admissible $\kappa,\rho$, a part of the space of Hamiltonians with a spectral gap is mapped continuously to a space of invertible finite-dimensional matrices. One can therefore hope that topological invariants that are stable under homotopies which do not close the spectral gap can be computed from homotopy-invariants of self-adjoint invertible matrices on the finite-dimensional side. And indeed, the spectral localizer method identifies those with index pairings of $H$ and $\DD$.

We describe this in more detail now for the complex pairings. For non-trivial pairings one always needs an additional symmetry: In the even-dimensional case the Dirac operator anti-commutes with a symmetry operator
$\gamma_0 \DD = - \DD \gamma_0$
while the Hamiltonian will not be assumed to have additional symmetry, while in the odd-dimensional case the Hamiltonian anti-commutes with a standard chiral symmetry $J H = -H J$ while the Dirac operator has no additional symmetry. In the latter case, we can write $$H=\begin{pmatrix} 0 & H_0^* \\ H_0 & 0\end{pmatrix}$$ as a matrix over $\Hh_+ \oplus \Hh_-$ which induces the unitary equivalence
$$L_\kappa \sim \begin{pmatrix} 0 & (L^{\mathrm{o}}_{\kappa})^*
	\\ L^{\mathrm{o}}_{\kappa} & 0\end{pmatrix}$$
where 
\begin{equation}
	\label{eq:oddlocalizer}L^{\mathrm{o}}_{\kappa} = \kappa\DD \otimes J + H = \begin{pmatrix}
		\kappa\DD & H_0^* \\ H_0 & - \kappa\DD
	\end{pmatrix}
\end{equation}
is the so-called odd spectral localizer. In the even case one can similarly decompose $L_\kappa$ into off-diagonal blocks with the even spectral localizer
\begin{equation}
	\label{eq:evenlocalizer}
	L^{\mathrm{e}}_{\kappa} = \kappa \DD + \gamma_0 \otimes H = \begin{pmatrix}
		H & \kappa D_0^* \\ \kappa D_0 & - H
	\end{pmatrix}
\end{equation}
where $D_0$ is the off-diagonal part of $\DD$ in its grading $\gamma_0$.

\begin{theorem}[{\cite{LS2,LS3}}]
\label{theorem:complexlocalizer}
Let $\DD$ be a $d$-dimensional Dirac operator, $H$ be a gapped short-range Hamiltonian and assume that $\symlabel \in \{e,o\}$ distinguishes between even and odd symmetry type where $d$ is even and $\DD$ has a chiral symmetry $\gamma_0$ respectively $d$ is odd and $H$ has a chiral symmetry $J$.

Then for each pair $\kappa$, $\rho$ such that
$$\kappa < \frac{g^3}{12 \norm{H}\, \norm{[\DD,H]}},\quad \rho > \frac{2g}{\kappa}$$
the finite-dimensional operator
$$L^\symlabel_{\kappa,\rho} :=\pi_\rho \circ L^\symlabel_\kappa \circ \pi_\rho^*$$ 
is a self-adjoint invertible finite-dimensional matrix with spectral gap 
$$\abs{L^\symlabel_{\kappa,\rho}} \geq \frac{g}{2} \pi_\rho \pi_\rho^*$$
and 
$$\Ind(F_{up}) = \frac{1}{2}\Sig(L^{\mathrm{e}}_{\kappa,\rho})$$
in the even case, respectively
$$\Ind(F_{pu}) = -\frac{1}{2}\Sig(L^{\mathrm{o}}_{\kappa,\rho})$$
in the odd case, where $\Sig$ denotes the number of positive minus the number of negative eigenvalues.
\end{theorem}

\vspace{.25cm}
The skew localizer is a generalization of this result to also allow the computation of $\ZM_2$-valued index pairings in terms of the signature, determinant or Pfaffian of a finite-dimensional matrix which is constructed from matrix elements of the Dirac operator and the Hamiltonian. The main result on the skew localizer for all Altland-Zirnbauer classes can be stated as follows:
\begin{theorem}[{\cite{DS21}}]
\label{theorem:skewlocalizer}
Let $\DD$ be a $d$-dimensional standard Dirac operator, and let $g > 0$ and $A > 0$ be some constants. Define the set
$$U^\sigma_{g,A} = \{H \in \mathrm{Ham}_{\symlabel}(n): \, \norm{H} \leq 1, \, \abs{H}\geq g \one, \,\norm{[\DD,H]}\leq A\}$$
where boundedness of the commutator shall also include the condition $H \Dom(\DD) \subset \Dom(\DD)$.

Let $(\kappa, \rho)$ be real numbers such that
$$\kappa < \frac{g^3}{12 A},\quad \rho > \frac{2g}{\kappa},$$
then there exists for each symmetry class $\sigma$ an $\RM$-affine linear map
$$L^\symlabel_{\kappa,\rho}: \mathrm{Ham}_{\symlabel}(n) \to M_m(\CM),$$ 
for some finite dimension $m\in \NM$ with the following properties:
\begin{enumerate}
\item[(i)] The map takes the form
$$H \mapsto (u^\sigma_\rho) \circ  L_{\kappa,\rho}(H) \circ (v_\rho^\sigma)^*$$  
for the finite-volume spectral localizer $L_{\kappa,\rho}(H)$ acting on $\pi_\rho(\Hh):=\pi_\rho(\CM^2\otimes \CM^{d'}\otimes \CM^{n} \otimes \Hh_0)$ and two partial isometries $u^\sigma_\rho,v^\sigma_\rho: \CM^m \to \pi_\rho(\Hh)$, depending on $n$, $\rho$, $\DD$ and the symmetry operators, but not on $H$ or $\kappa$. 
\item[(ii)]
For each $H \in U^\sigma_{g,A}$ the image is an invertible matrix with spectral gap
$$\abs{L^\symlabel_{\kappa, \rho}(H)} \geq \frac{g}{2} \one_m.$$
\item[(iii)] Depending on $\symlabel$, $U^\sigma_{g,A}$ is mapped under $L^\sigma_{\kappa,\rho}$ either to the self-adjoint, real self-adjoint or real skew-adjoint matrices which are invertible and there is on the range a locally constant map $\Sig_\symlabel$ which takes values in $\ZM$ or $\ZM_2$ and is up to a sign given by half of the signature, sign of the determinant respectively the sign of the Pfaffian.
\item[(iv)] For each $H \in U^\sigma_{g,A}$ one has
$$\Ind_\symlabel(F) = \Sig_\symlabel(L^\symlabel_{\kappa,\rho}(H))$$
for $F \in \{F_{pp},F_{pu}, F_{up}\}$ the Fredholm operator constructed from $H$ and $\DD$ appropriate for the symmetry class $\sigma$. 
\end{enumerate}

\end{theorem}
\noindent{\bf Proof.}
The skew-localizers for all non-trivial cases $(j,d)$ are constructed in \cite{DS21} with slightly different notation since the symmetry relations are imposed upon the off-diagonal parts $H_0$, $D_0$ whenever there is a chiral symmetry. The form of the symmetry operators in Definition~\ref{def:standard_symmetries} respectively the symmetry operators induced by the choice of Clifford representation for Definition~\ref{def:standarddirac} is compatible with all necessary algebraic conditions.
\hfill $\Box$

Thus, one can construct a matrix $L^\symlabel_{\kappa,\rho}(H)$ from the finite-dimensional matrices $\pi_\rho H \pi_\rho^*$ and $\pi_\rho \DD \pi_\rho^*$, which is invertible if $H$ is invertible and the parameters $\kappa$, $\rho$ are chosen appropriately. Each of the functions $\Sig_\symlabel$ (which implicitly depends on $\rho$) is constant on each connected component of the invertible matrices with the type of symmetry resulting from the combined symmetries of $\DD$ and $H$. For the odd/even pairings $L^{o}_{\kappa,\rho}, L^e_{\kappa,\rho}$ given by the expressions \eqref{eq:oddlocalizer},\eqref{eq:evenlocalizer} it is easy to see that the dependence on $H$ and $\DD$ has the stated form. The precise form of $L^\symlabel_{\kappa,\rho}$ for the $\bbZ_2$-valued pairings depends on the symmetry class and the symmetry operators. To construct it, one starts with the self-adjoint form of the spectral localizer \eqref{eq:sl_big} respectively its finite-volume restrictions, which can in every symmetry class be brought into a more specialized block-diagonal form via a symmetry-adapted change of basis. The skew-localizer $L^\symlabel_{\kappa,\rho}$ is then a specific block of that matrix.

\section{Continuity of the index pairing for random families}

Consider a norm-continuous family of Hamiltonians $t\in T\mapsto H(t)$ labeled by a topological space such that each $H(t)$ has a well-defined index pairing with respect to some fixed Dirac operator. Those index pairings depend only on the band-flattened Hamiltonians $t \in T\mapsto \sgn(H(t))$, which is also a norm-continuous family if there is a uniform spectral gap, i.e. an interval $\Delta$ containing $0$ such that $\Delta\cap \mathrm{Spec}(H(t))=\emptyset$.
In our applications we will, however, want to also consider Hamiltonians which have dense pure-point spectrum in an interval containing $0$ and therefore one cannot expect the band-flattenings of the Hamiltonians to have a norm-continuous parameter-dependence. 

Let us recall briefly recall the basic problem here. The topological invariants in that mobility-gapped regime are still given by index pairings, i.e. the indices of Fredholm operators. It is well-known that the index is constant on norm-continuous paths of Fredholm operators. However, this topology is too strong to apply to perturbations in the mobility gap regime. This leads to the slightly paradoxical situation that even though our topological invariants are apparently protected by index theorems, a priori nothing prevents the indices from changing under tiny perturbations of the Hamiltonians.

In the complex symmetry classes one can prove better invariance results by using local formulas for the indices, which then only require forms of weak operator continuity to prove that the indices change continuously. The constancy of the index under perturbations is then a simple consequence of its discreteness. We can use the spectral localizer as a substitute for the local index formulas, since the reduction to a finite-dimensional subspace automatically improves weak operator continuity of families to norm-continuity. The remaining problem is that for a fixed realization of a random disordered system one does not even have weak continuity for the band-flattenings of the Hamiltonians, hence we can at most ask for some averaged form of weak continuity.

In this section we implement such a continuity argument in an abstract setting before we later prove that its assumptions are satisfied in a large class of relevant models. Throughout we assume that there is a fixed Dirac operator $\DD$ as in Definition~\ref{def:standarddirac}.

\begin{definition}
\label{def:randomfamily}
Let $(H_\omega)_{\omega\in \Omega}$ be a measurable family of self-adjoint operators acting on $\Hh_0\otimes \CM^n$ for $(\Omega, \PP)$ consisting of a compact Hausdorff space $\Omega$ with a regular Borel measure $\PP$.

We say that $H$ is a random family of gapped short-range Hamiltonians if the following holds:
\begin{enumerate}
	\item[(i)] The family is essentially bounded $\esssup_{\omega\in \Omega} \norm{H_\omega} < M < \infty$.
	\item[(ii)] There exists an open interval $\Delta$ including $0$ such that almost surely $\Delta \cap \mathrm{Spec}(H_\omega)=\emptyset$.
	\item[(iii)] We have almost surely $H_\omega  \Dom(\DD) \subset \Dom(\DD)$ and therefore the commutator $[\DD,H_\omega]$ is well-defined and symmetric on $\Dom(\DD)$.
	
	\item[(iv)]  There is a function $f \in L^1(\Omega)$ such that
	\begin{align*}
		\|[\DD,H_\omega]\| &\leq f(\omega)
	\end{align*}
	in the sense that this operator almost surely extends from $\Dom(\DD)$ to a bounded operator which satisfies the stated norm bound.
\end{enumerate}
\end{definition}

In our applications the family $H_\omega$ will arise from some random family of physical Hamiltonians by spectral flattening $H_\omega=\sgn(h_\omega)$ and therefore consists of invertible self-adjoint operators with spectral gap $\Delta=(-1,1)$. 

\begin{definition}
\label{def:contonaverage}
Let $T$ be a topological space. A map $t\in T \mapsto (H_\omega(t))_{\omega \in \Omega}$ will be called continuous on average if 
\begin{enumerate}
    \item[(i)] For each $t\in T$ the family $(H_\omega(t))_{\omega \in \Omega}$ is a random family of gapped short-range Hamiltonians as in Definition~\ref{def:randomfamily} with norm bounds $M(t)$ and $L^1(\Omega)$-functions $f(t)$.
    \item[(ii)] The interval $\Delta$ can be chosen independently of $t$, the constants $M(t)$, and the norms $\norm{f(t)}_{L^1(\Omega)}$ are locally bounded for each $t$.
    \item[(iii)] The direct integral operators
$$t\in T \mapsto \int_{\Omega}^\oplus H_\omega(t) \bbP(\difd{\omega})$$
are continuous in the weak operator topology of $L^2(\Omega)\otimes \Hh_0 \otimes \CM^n$. 
\end{enumerate}
\end{definition}
A consequence of the weak operator continuity is that for every $\psi_1,\psi_2 \in  \Hh_0 \otimes \CM^n$ the matrix element
$$t \in T \mapsto \langle \psi_1, H_\omega(t) \psi_2\rangle$$ as a random variable is continuous in the norm of $L^1(\Omega)$ (applying weak operator continuity to the vectors $\one_\Omega\otimes \psi_1, \one_\Omega\otimes \psi_2$). Explicitly, the averaged differences satisfy
\begin{equation}
\label{eq:convergence_average_differences}
\lim_{t'\to t}\int_{\Omega} \abs{\langle \psi_1| \left(H_\omega(t)-H_\omega(t')\right)\psi_2\rangle_{ \Hh_0 \otimes \CM^n}}\bbP(\difd{\omega}), \quad \forall t\in T, \psi_1,\psi_2 \in \Hh_0 \otimes \CM^n.
\end{equation}
In the following we attach to all derived operators such as the spectral localizer a subscript $\omega$ respectively the argument $t$ to specify the underlying $H_\omega(t)$. The true index of a configuration is $\Ind(\omega,t) := \Ind_{\symlabel}(F_{\omega}(t))$ where $F_\omega$ is a Fredholm operator depending on $\DD$, $H_\omega$ and the symmetry class $\symlabel$; we can probe it numerically using the finite-volume signature $\Ind_{\kappa,\rho}(\omega,t):=\Sig_{\symlabel}(L^{\symlabel}_{\kappa,\rho,\omega}(t))$ for suitable $\kappa$, $\rho$. 

For any $\epsilon>0$ we can choose $\kappa$ so small and $\rho$ so large that the finite volume signature coincides with the true index up to a probability $\epsilon$. The idea is now to use continuity on average to show that those finite volume signatures are locally constant with large probability. The conclusion will be that the distribution of the true index depends continuously on the parameter $t\in T$ (though it is not necessarily constant, since there can at each stage of the argument be rare enough events where the index could possibly jump).

\begin{proposition}
\label{prop:continuity}
Let $t\in T \mapsto H(t)$ be a parametrized random family of gapped short-range Hamiltonians which is continuous on average and almost surely takes values in some fixed $\mathrm{Ham}_{\symlabel}(n)$ as in Definition~\ref{def:symmetries}. Then there is for every $\epsilon>0$ and each $t_0 \in T$ a neighborhood $U$ of $t_0$ such that 
$$\bbP(\Ind(\omega,t)=\Ind(\omega,t_0)) \geq 1-\epsilon, \qquad \forall t,t_0\in U.$$
Equivalently, $t \in T \mapsto \bbP(\Ind(\omega,t)=z)$ is continuous for each $z\in \bbZ$ respectively $z\in \ZM_2$.
\end{proposition}
\noindent{\bf Proof.}
We will in the following assume without loss of generality that $M(t)< 1$ and $\Delta = [-g,g]$ for some $g>0$. 

Choose a neighborhood $V$ of $t_0$ such that $\norm{f(t)}_{L^1(\Omega)} < C < \infty$ holds for some $C>0$ and all $t\in V$. One has by the Markov inequality
$$\bbP(\norm{[\DD,H_\omega(t)]} > \frac{2C}{\epsilon}) \leq \frac{\epsilon}{2}$$
for all $t \in V$. For the given $\epsilon>0$ choose any
$$\kappa < \epsilon\frac{g^3}{24 C}, \quad \rho > \frac{2g}{\kappa}$$
which implies
\begin{equation}
	\label{eq:tech1}
	\bbP(\Ind(\omega,t)=\Ind_{\kappa,\rho}(\omega,t)) \geq 1-\frac{\epsilon}{2}
\end{equation}
for all $t\in V$ since the sufficient condition for Theorem~\ref{theorem:skewlocalizer} is satisfied on a set of probability greater than $1-\frac{\epsilon}{2}$.

From Theorem~\ref{theorem:skewlocalizer}(i) it is clear that the weak operator continuity of $t\in T \mapsto H(t)$ implies weak operator continuity of $t \in T\mapsto \int_\Omega^\oplus L^\symlabel_{\kappa,\rho,\omega}(t)\difd\PM(\omega)$. Since that implies convergence of $L^1(\Omega)$-averaged matrix elements as in \eqref{eq:convergence_average_differences} and the localizer is finite-dimensional one can therefore choose for the given $\epsilon$, $\kappa$, $\rho$ a smaller neighborhood $U \subset V$ such that $$\norm{L^{\symlabel}_{\kappa,\rho}(t)-L^{\symlabel}_{\kappa,\rho}(t_0)}_{L^1(\Omega, M_m(\bbC))} \leq \frac{1}{8}g \epsilon$$
for all $t\in U$, where $M_m(\CM)$ carries the operator norm and $m$ is the matrix size of the spectral localizer (which depends on $\rho$).

By the Markov inequality one has then 
$$\mathbb{P}(\norm{L^\symlabel_{\kappa,\rho,\omega}(t)-L^\symlabel_{\kappa,\rho,\omega}(t_0)}_{M_m(\bbC)} > \frac{g}{4}) \leq \frac{\epsilon}{2}.
$$
Each Hamiltonian has a spectral gap in $[-g,g]$ and the associated spectral localizer therefore has a spectral gap in at least $(-\frac{g}{2},\frac{g}{2})$ assuming $\kappa,\rho$ are admissible. Consider the path 
$$\lambda \in [0,1] \mapsto L(\lambda):=(1-\lambda) L^\symlabel_{\kappa,\rho,\omega}(t) + \lambda L^\symlabel_{\kappa,\rho,\omega}(t_0),$$ which linearly interpolates between the localizers of $H_{\omega}(t)$ and $H_{\omega}(t_0)$ and preserves whichever (real) symmetries the localizer has. Due to
$$\norm{L(\lambda)-L(0)}\leq \lambda \norm{L^{\symlabel}_{\kappa,\rho,\omega}(t)-L^{\symlabel}_{\kappa,\rho,\omega}(t_0)}$$
the spectral gap of $L(\lambda)$ is at least $(-\frac{g}{4},\frac{g}{4})$ whenever the right-hand side of this equation is less than $\frac{g}{4}$.

There is therefore an intersection of two measurable subsets of $\Omega$, both of which have probability greater than $1-\frac{\epsilon}{2}$, on which $L^\symlabel_{\kappa,\rho}(t)$ and $L^\symlabel_{\kappa,\rho}(t')$ are connected by a symmetry-preserving continuous homotopy within the invertible matrices, hence have the same signature $\Sig_\symlabel$. We conclude that for all $t\in U$ one has
\begin{align*}
&\bbP(\Ind(\omega,t)=\Ind(\omega,t_0))\\
&\geq \bbP(\Ind(\omega,t)=\Ind_{\kappa,\rho}(\omega,t)=\Ind_{\kappa,\rho}(\omega,t_0)=\Ind(\omega,t_0))\\
&\geq 1- \epsilon.
\end{align*}
Since the index takes only discrete values this shows continuity of the probability distribution.
\hfill $\Box$

In particular, if the index almost surely does not depend on $\omega$, as is the case e.g. for ergodic quantum systems, then continuity implies the following:
\begin{corollary}
\label{cor:almost_surely_constant}
In the situation of the Proposition, if for each fixed $t$ there is some $z(t)$ with $\bbP\left(\Ind(\omega,t)=z(t)\right)=1$ then $z(t)$ is locally constant.
\end{corollary}

\section{Mobility-gapped Hamiltonians}
\label{sec:mbg}
We will consider tight-binding Hamiltonians on a discrete $d$-dimensional Delone pattern $\Ll$, i.e. a set $\Ll\subset \RM^d$ which is uniformly discrete and relatively dense. Hamiltonians are bounded operators in $\Bb(\ell^2(\Ll)\otimes \CM^n)$ and Dirac operators act on $\ell^2(\Ll)\otimes \CM^{d'}$, whenever necessary we consider them to be amplified with trivial tensor products to make them act on a common Hilbert space with fiber dimension $N=nd'$. We denote for any $x\in \Ll$ by $\langle x|$ respectively $|x\rangle$ the obvious partial isometry $\ell^2(\Ll)\otimes \CM^N \to \CM^N$ respectively its adjoint. Any operator $O\in \Bb(\ell^2(\Ll)\otimes \CM^N)$ is therefore defined by its matrix elements $\langle x| O|y\rangle \in M_N(\CM)$. We fix throughout some Altland-Zirnbauer class $\symlabel$ (either complex or real) and with the usual complex conjugation of $\ell^2(\Ll)$ we assume that time-reversal and particle-hole conjugation are implemented anti-unitarily by standard symmetry operators as in Definition~\ref{def:standard_symmetries}, while any chiral symmetry is also standard as in Definition~\ref{def:chiralsymmetry}. 

Choose an arbitrary offset $z=\sum_{j=1}^d z_j e_j \in \RM^d\setminus \Ll$ and define the standard Dirac operator
$$\DD = \sum_{j=1}^d \gamma_j \otimes (X_j- z_j)$$
where $\gamma_1,...,\gamma_d$ is the irreducible representation of the Clifford algebra of Appendix~\ref{sec:app} and the position operators $X_1,...,X_d$ act by multiplication $X_j|x\rangle= x_j |x\rangle$ for all $x\in \Ll$. The square of $\DD$ is $\sum_{j=1}^d (X_j-z_j)^2$ which has a strictly positive lower bound by the choice of $z$. It is easily seen that $\DD$ has purely discrete spectrum and compact resolvent, in fact $\abs{\DD}^{-1}$ is $p$-Schatten class for all $p > d$.

Let us now describe a concrete example to illustrate the index pairings and the skew localizer method. We consider a Hamiltonian in two dimensions with four on-site degrees of freedom, i.e. a self-adjoint operator $h$ on $\ell^2(\ZM^2) \otimes \CM^4$. We assume that $h$ is in the AII symmetry class, which means that the time-reversal operator should satisfy $\mathfrak{T}^2=-\one$ and $\mathfrak{T}h\mathfrak{T}^*=h$. A simple example \cite{HughesPRB2011} for an AII-class topological insulator is
$$h= \frac{1}{2\imath}(S_1-S_1^*)(\sigma_3\otimes\sigma_1)+\frac{1}{2\imath}(S_2-S_2^*)(\one_2\otimes\sigma_2)+ (m - \frac{1}{2}(S_1+S_1^*+S_2+S_2^*))(\one_2\otimes \sigma_3)$$
with the directional shifts $S_1,S_2$ on $\ell^2(\ZM^2)$ and the time-reversal operator $\mathfrak{T}=\imath(\sigma_2\otimes\one_2) \Cc$. This corresponds to the case $j=4$ in Definition~\ref{def:symmetries} and, indeed, $\mathfrak{T}$ is in the standard form Definition~\ref{def:standard_symmetries}.

The Hamiltonian is short-range (due to $\norm{[\DD,S_i]}\leq 1$) and gapped if $m\notin \{-2,0,2\}$ as one can check by Fourier-transforming it to a matrix-valued multiplication operator on $L^2(\TM^2)\otimes \CM^4$. Its spectral projection $P=\chi(h<0)$ defines an element of the Real $K$-theory group $\mathrm{KR}_4(C(\TM^2))$ and it is non-trivial in the range $-2<m<2$, as signified by its Kane-Mele $\ZM_2$-invariant, which corresponds to the strong topological invariant.

In two dimensions the Dirac operator takes the form
$$\DD=\begin{pmatrix}
	0 & (X_1-z_1)+\imath (X_2-z_2) \\ (X_1-z_1) - \imath (X_2-z_2) & 0
\end{pmatrix},$$
with the matrix decomposition w.r.t. to another set of on-site degrees of freedom separate from those of $h$, i.e. the spectral localizer
$$L_\kappa = \begin{pmatrix}
	\one_2\otimes h  & \kappa \DD\otimes \one_4\\ \kappa \DD\otimes \one_4 & -\one_2 \otimes h
\end{pmatrix}$$
acts on $\CM^2\otimes \CM^2\otimes \CM^4\otimes \ell^2(\ZM^2)$. The Dirac operator is chiral and the off-diagonal part of the Dirac operator has the real symmetry $\overline{D_0} = \Cc D_0 \Cc= D_0^*$, i.e. $D_0=D_0^t$. 

The relevant index pairing is $\Ind_2(F_{up})$ with $F_{up}$ given by \eqref{eq:freds1}. To construct the skew localizer that computes it, one first reduces to the odd localizer
$$L_\kappa^o(h)=\begin{pmatrix}
	h  & \kappa D_0^*\otimes \one_4\\ \kappa D_0\otimes \one_4 & - h
\end{pmatrix}$$
which is an off-diagonal block of $L_\kappa$. Due to the symmetries of $h$ and $\DD$ the odd localizer has the real symmetry $$Q^*\overline{L_\kappa^o}Q=-L_\kappa^o, \qquad Q = \begin{pmatrix}
	0 & u \\ -u & 0
\end{pmatrix}$$
where $u$ is the real self-adjoint unitary such that $\mathfrak{T}=u \Cc$.

To construct the skew-localizer according to \cite{DS21} one then needs to find a unitary square-root $R=R^t$ with $R^2=Q$. Such a $4\times 4$-matrix always exists and can be constructed via diagonalization of $Q$. Then one sets finally
$$L^{AII}_\kappa(h) = \imath R L_\kappa R^*, \quad L^{AII}_{\kappa,\rho}= \pi_\rho\circ L^{AII}_{\kappa} \circ \pi_\rho^*,$$
which is real skew-adjoint (i.e. equal to its complex conjugate and to the negative of its adjoint). For an explicit  expression that applies to the present special case we refer to \cite{SB24}. The finite-volume restrictions $L^{AII}_{\kappa,\rho}$ are finite-dimensional block matrices expressed in terms of the finitely many matrix elements $\langle x|  D_0 |y\rangle$ and $\langle x| h | y\rangle$ for $x,y\in \ZM^2$ with modulus less than $\rho$. In real-space we therefore only need the Hamiltonian in a finite ball around the origin to compute the spectral localizer, which makes the expression extremely local.

Choosing first $\kappa>0$ small enough and then $\rho$ large enough, the $\ZM_2$-valued index can be computed as $$\Ind_2(P \frac{X_1+\imath X_2-z}{\abs{X_1+\imath X_2-z}} P + 1-P)= \sgn(\mathrm{Pf}(L^{AII}_{\kappa,\rho}(h))\sgn(\mathrm{Pf}(L^{AII}_{\kappa,\rho}(\tilde{h})))\in \{-1,1\}$$
where $P$ is the Fermi projection of $h$, $\mathrm{Pf}$ is the Pfaffian and one multiplies with the skew localizer of a topologically trivial Hamiltonian in the same symmetry class as $h$, for example $\tilde{h}=\one_{\ell^2(\Ll)}\otimes \one_2\otimes \sigma_3$ to make sure  that a value of $-1$ corresponds to the topologically non-trivial phase and that the right-hand side does not depend on the choice of basis used to compute the Pfaffian.

We can now include some disorder in the model by promoting the Hamiltonian to a random family $(h_\omega)_{\omega\in \Omega}$. For example, we can use the product space $\Omega= [0,1]^{\ZM^2}$ and have Anderson-type i.i.d. on-site disorder
\begin{equation}
\label{eq:example_ham_disorder} 
h_\omega = h + \lambda \sum_{x\in \ZM^2} \omega_x \one_4 \otimes |x\rangle\langle x|
\end{equation}
which is still time-reversal symmetric. If $h$ is gapped and the disorder parameter $\abs{\lambda}$ is small each $h_\omega$ will still be a gapped Hamiltonian independent of $\omega$, but for large enough disorder the gap will close. One would then expect to be in a mobility-gapped phase, i.e. for almost every $\omega$ one has pure point spectrum in a spectral region around $0$ with exponentially localized eigenfunctions.   The results of \cite{DdNSb} can be used to show that Hamiltonians like \eqref{eq:example_ham_disorder} have a mobility gap in the sense of Definition~\ref{def:mbg} for a non-trivial spectral interval $\Delta$ even when the disorder parameter $\lambda$ is large enough to close the spectral gap and also for more general forms of disorder which are off-diagonal (as would be required to satisfy a particle-hole symmetry). The main result of this paper implies that one will then still have almost surely the same $\ZM_2$-index as in the clean limit $\lambda=0$.

The remainder of this section will be used to show that the band-flattenings of mobility-gapped Hamiltonians determine random families of operators as in Definition~\ref{def:randomfamily} and which are in the parametrized case continuous on average as in Definition~\ref{def:contonaverage}. This constitutes the proof of Theorem~\ref{th:main}.

The following is well-known (e.g. \cite{Prodan2017,PSbook}), but we repeat the proof since we will need to use the same strategy again later.
\begin{lemma}
\label{lemma:fermiprojectiondecay}
If $h=(h_\omega)_{\omega\in \Omega}$ is an essentially bounded measurable family with a mobility gap $\Delta$ containing $0$ and $0$ itself is almost surely not an eigenvalue of $h_\omega$ then the flattened Hamiltonian $H_\omega=\sgn(h_\omega)$ satisfies for any $k\in \bbN$ a bound
$$\sup_{x,y\in \Ll} \langle x-y\rangle^{k} \int_\Omega \norm{\langle x|H_\omega|y\rangle} \,\difd{\bbP(\omega)} \leq \tilde{A}_k $$
with $\tilde{A}_k$ depending only on the constants of the mobility gap $A_{k}(\delta)$ for some $\delta > 0$ and on $\norm{h}$.
\end{lemma}
\noindent{\bf Proof.} Consider a counter clockwise rectangular loop $\Cc$ in the complex plane enclosing the interval $[-\norm{h}, 0]$ which meets the real line in $0$ and satisfies $\mathrm{dist}(\Cc, \mathrm{Spec}(h)\setminus \Delta)>\delta$ for some $\delta>0$. The intersection of $\Cc$ with the region $\abs{\mathrm{Im}(\eta)}>\epsilon$ consists of two disconnected curves $\Cc^{\pm}_\epsilon$. With those regularized curves one can write
$$\frac{1}{2}(1-\sgn(h_\omega)) = -\frac{1}{2\pi} \slim_{\epsilon\to 0} \sum_{s\in \{-,+\}}\int_{\Cc^{s}_\epsilon} \frac{1}{h_\omega-\eta} \difd \eta$$
where the convergence in the strong operator topology needs that $0$ is not an eigenvalue. For $x\neq y$ we can therefore bound
$$\norm{\langle x| H_\omega |y\rangle}  \leq 
\frac{1}{\pi}\int_{\Cc} \int_\Omega\lVert\langle x|\frac{1}{h_\omega-\eta}|y\rangle\rVert \difd{\bbP(\omega)} \difd \eta.$$
Due to the mobility gap and the standard resolvent estimate $\lVert\langle x|(h_\omega-\eta)^{-1}|y\rangle|\rVert\leq  \mathrm{Im}(\eta)^{-1}$ one has
\begin{align*}
\int_{\Cc} \int_\Omega\lVert\langle x|\frac{1}{h_\omega-\eta}|y\rangle\rVert \difd{\bbP(\omega)} \difd \eta &\leq \int_{\Cc}  \mathrm{Im}(\eta)^{1-\nu}\int_\Omega\lVert\langle x|\frac{1}{h_\omega-\eta}|y\rangle\rVert^\nu \difd{\bbP(\omega)} \difd \eta\\
&\leq A_k(\delta)  \langle x-y\rangle^{-k}\int_{\Cc} \mathrm{Im}(\eta)^{1-\nu} \difd \eta
\end{align*}
for any $k>0$ and $0 < \nu < 1$ which yields a convergent integral since the singularity at the real line is integrable. Note that one can always choose $\Cc$ in such a way that the length of $\Cc$ is bounded by a universal constant times $(1+\norm{h})$.
\hfill $\Box$

This immediately implies a useful deterministic bound:
\begin{lemma}
\label{lemma:det_decay}
Under the conditions of  Lemma~\ref{lemma:fermiprojectiondecay} there is for each $k\in \bbN$ a function $C_k \in L^1(\Omega)$ such that almost surely
	\begin{equation}
	    \label{eq:simple_me_estimate}
	\norm{\langle x| H_\omega|y\rangle} \leq C_k(\omega) {\langle y\rangle^{d+1}}{\langle x-y \rangle^{-k}}.
 \end{equation}
\end{lemma}
\noindent{\bf Proof.}
For the positive measurable function $$C_k(\omega):= \sum_{x,y\in \Ll} \langle y\rangle^{-(d+1)}\langle x-y\rangle^k\norm{\langle x| H_\omega| y \rangle}$$ one has finite expectation value $\mathbb{E}(C_k) < \infty$ whenever $k$ is large enough.
\hfill $\Box$

The Dirac operator has the core $\Ee = \cap_{k\in \NM} \mathrm{Dom}(\DD^k)$ consisting of precisely those functions on $\Ll$ which decay faster than any inverse polynomial. Since for any fixed $y\in \Ll$ the matrix elements $\langle x|H_\omega |y\rangle$ almost surely decay faster than any inverse polynomial in $x$, one sees that $H_\omega$ almost surely maps $\Ee$ into itself. The commutator $[\DD,H_\omega]$ is therefore almost surely well-defined on that dense subset, however, we do not expect our localization assumptions to imply almost sure boundedness of it. We can at most prove that this operator in some sense does grow only very slowly at infinity:
\begin{lemma}
\label{lemma:comm_est_tech}
Under the conditions of Lemma~\ref{lemma:fermiprojectiondecay} there is for any $s>0$ a positive function $f_s\in L^1(\Omega)$ such that 
	\begin{equation}
	\label{eq:weak_comm_bound}
 \norm{[\DD, H_\omega]w^{s}} \leq f_s(\omega)
        \end{equation}
        with $w=(1+\DD^2)^{-\frac{1}{2}}$. Moreover, the $L^1$-norm of $f_s$ depends only on the constants $\tilde{A}_k$ of Lemma~\ref{lemma:fermiprojectiondecay}.
\end{lemma}
\noindent{\bf Proof.}
Introduce the projections $P_x=|x\rangle \langle x|$ and $P_R=\sum_{\abs{x}<R} P_x$ which converge strongly to $\one$ for $R\to \infty$. Note that $P_R[\DD,H_\omega]P_R$ is a bounded operator for every $R$ and one can conclude from Lemma~\ref{lemma:det_decay} that for almost every $\omega\in \Omega$ one has for every $\psi \in \Ee$ strong convergence $$\lim_{R\to\infty}P_R[\DD,H_\omega]w^s P_R \psi = [\DD,H_\omega]w^s \psi.$$
The $C^*$-identity implies that for any $q\in \NM$ one has
$\norm{a}=\norm{(a^*a)^q}^{1/(2q)}$ which leads to
\begin{align}
\norm{P_R [\DD,H_\omega]w^{s} P_R}^{2q}&=\lVert \prod_{i=1}^q (P_R w^{s}[\DD,H_\omega] P_R [\DD,H_\omega] w^{s}P_R)\rVert \nonumber\\
&=\lVert \sum_{x_1,...,x_{2q+1} \in \Ll} \prod_{i=1}^q  (P_{x_{2i-1}} P_R w^{s}[\DD,H_\omega]P_R P_{x_{2i}} P_R [\DD,H_\omega]w^{s}P_R P_{x_{2i+1}}) \rVert \nonumber \\
&\leq \sum_{x_1,...,x_{2q+1} \in \Ll} \prod_{i=1}^q  \lVert P_{x_{2i-1}} w^{s}[\DD,H_\omega]P_{x_{2i}} \rVert \lVert P_{x_{2i}} [\DD,H_\omega]w^{s} P_{x_{2i+1}} \rVert \label{eq:tech_est1}
\end{align} 
All sums are finitely supported except for the last one. The final right hand side does not depend on $R$ anymore; if it is finite then $\lVert P_R [\DD,H_\omega]w^{s}P_R\rVert$ is bounded uniformly in $R$ and since the strong operator limit for $R\to \infty$ exists at least on the dense subspace $\Ee$ the limit satisfies the same norm bound and must then be the closure of the a priori only densely defined operator $[\DD,H_\omega]w^{s}$.

By definition of $P_x$, $\DD$ and $w$ one has
\begin{align*}
	\norm{P_x w^{s} [\DD,H_\omega] P_y} &=  \langle x-z\rangle^{-s} \norm{(x\cdot \gamma -y\cdot \gamma)P_x H P_y} \\
	&\leq c\langle{x}\rangle^{-s} \abs{x-y} \norm{P_x H_\omega P_y}
\end{align*}
with some offset $z\in \RM^d\setminus \Ll$ and where both sides are still random variables depending on $\omega\in \Omega$. To show that \eqref{eq:tech_est1} is almost surely finite it is enough to show that its $L^1(\Omega)$-norm is finite. Therefore we take the expectation value of \eqref{eq:tech_est1} then apply the H\"older inequality of $L^p(\Omega)$ to split into $2q$ factors, each of which can be estimated using
$$\left(\int_{\Omega} \norm{P_x H_\omega P_y}^{2q} \mathrm{d}\PM(\omega)\right)^{1/(2q)} \leq \left(\int_{\Omega} \norm{P_x H_\omega P_y} \mathrm{d}\PM(\omega)\right)^{1/(2q)} \leq \tilde{A}_{2qk}^{1/(2q)} \langle x-y\rangle^{-k}$$
where we used that the operator norm $\norm{P_x H_\omega P_y}$ is less or equal $1$ and then the estimate of Lemma~\ref{lemma:fermiprojectiondecay}. 

Thus there exists for any $k$ a constant such that the expectation value of \eqref{eq:tech_est1} can be bounded as
\begin{align*}\int_{\Omega} &\sum_{x_1,...,x_{2q+1} \in \Ll} \prod_{i=1}^q \lVert P_{x_{2i-1}} w^{s}[\DD,H_\omega]P_{x_{2i}} \rVert \lVert P_{x_{2i}} [\DD,H_\omega]w^{s} P_{x_{2i+1}} \rVert \\
\leq C_k &\sum_{x_1,...,x_{2q+1} \in \Ll}  \prod_{i=1}^{q} \langle x_{2i-1}\rangle^{-s} \langle x_{2i}-x_{2i-1}\rangle^{-k}  \langle x_{2i+1}-x_{2i}\rangle^{-k}\langle x_{2i+1}\rangle^{-s}
\end{align*}
with $C_k$ depending only on the constants $\tilde{A}_k$.

It remains to show that the expression is finite if one picks for given $s>0$ first $q$ and then $k$ large enough. For that we note the elementary estimate
\begin{align*}
	\sum_{x \in \Ll}  \langle x\rangle^{-\xi} \langle x-y\rangle^{-k} &\leq
	\sum_{|x|> \frac{\abs{y}}{2}} \langle 2 y\rangle ^{-\xi} \langle x-y\rangle^{-k} + \sum_{|x|\leq \frac{\abs{y}}{2}}  \langle x\rangle^{-\xi} \langle\frac{1}{2}y \rangle^{-k}\\
	&\leq c_{k,\xi} \left(\langle y\rangle^{-\xi}  + \langle y \rangle^{d-k}\right) \leq \tilde{c}_{\xi,k} \langle y\rangle^{-\xi}
\end{align*}
which holds for any $k > d + \max(\xi,1)$ with constants only depending on $\xi\geq 0$ and $k$. Applying it two times one has
\begin{align*}
&\sum_{x_{2q},x_{2q+1}\in \Ll} \langle x_{2q-1}\rangle^{-s} \langle x_{2q}-x_{2q-1}\rangle^{-k} \langle x_{2q+1}-x_{2q}\rangle^{-k} \langle x_{2q+1}\rangle^{-s} \\
&\leq \tilde{c}_{s,k} \sum_{x_{2q}\in \Ll} \langle x_{2q-1}\rangle^{-s} \langle x_{2q}-x_{2q-1}\rangle^{-k} \langle x_{2q}\rangle^{-s}
\leq \tilde{c}^2_{s,k} \langle x_{2q-1}\rangle^{-2s}.
\end{align*}
Iterating it another $2n-2$ times one ends up with 
$$\sum_{x_1,...,x_{2q+1} \in \Ll}  \prod_{i=1}^{q} \langle x_{2i-1}\rangle^{-s} \langle x_{2i}-x_{2i-1}\rangle^{-k} \langle x_{2i+1}-x_{2i}\rangle^{-k} \langle x_{2i+1}\rangle^{-s} \leq c \sum_{x_{1}\in \Ll} \langle x_{1}\rangle^{-2qs}
$$
which requires $k> d+\max(2qs,1)$. The final sum is finite whenever $2qs > d$ hence we can first choose $q$ that large and then any $k>d+\max(2qs,1)$ to complete the proof.
\hfill $\Box$

We can now apply a regularization trick that is motivated by similar problems in the context of spectral triples: We rescale the Dirac operator $\DD$ to $\DD^{(r)}=\DD (1+\DD^2)^{-r/2}$ for some fractional exponent $0 < r < 1$ to improve the regularity of the commutator. As we show in Proposition~\ref{prop:regularisation} of Appendix~\ref{sec:app2} the following is a consequence of Lemma~\ref{lemma:comm_est_tech}:
\begin{corollary}
\label{cor:rescaled_dirac_commutator}
Under the conditions of Lemma~\ref{lemma:comm_est_tech} there exists for any $0<s<r<1$ a constant $c_{r,s}$ depending neither on $H$ nor $\DD$ such that
$$\norm{[\DD^{(r)},H_\omega]}\leq c_{r,s}f_s(\omega).$$
\end{corollary}
Thus $H=(H_\omega)_{\omega\in\Omega}$ is a random family satisfying the properties of Definition~\ref{def:randomfamily} w.r.t. a properly rescaled Dirac operator. Note that the rescaling does not affect the symmetries of the Dirac operator nor the index pairings (the latter depend only on the flattening $\sgn(\DD)$ by the definition of the Fredholm operators \eqref{eq:freds1} and \eqref{eq:freds2}).

It now remains to prove that continuous families of Hamiltonians $t\in T \mapsto h_\omega(t)$ give after band-flattening rise to families $t\in T \mapsto H_\omega(t)$ that are continuous on average as in Definition~\ref{def:contonaverage}:

\begin{lemma}
\label{lemma:contmbgfamily}
Let $T$ be a metric space and for each $\omega \in \Omega$ let $t\in T \mapsto h_\omega(t)$ be maps such that each $(h_\omega(t))_{\omega\in \Omega}$ is a measurable family of bounded self-adjoint operators on $\ell^2(\Ll)\otimes \CM^n$.
Assume
\begin{enumerate}
\item[(i)] uniformly short hopping range in the sense that for each $k\in \bbN$
$$\sup_{t\in T,\omega\in\Omega,x,y\in \Ll}\langle x-y\rangle^{k}\norm{\langle x| h_\omega(t) | y \rangle}_{M_n(\bbC)} <\infty.$$
\item[(ii)] averaged continuity of the matrix elements in the sense that
$$t\in T \mapsto \langle x| h(t)| y\rangle \in L^2(\Omega,M_n(\CM))$$ is continuous for each $x,y\in \Ll$.
\item[(iii)] a uniform mobility gap, {\sl i.e.} the set $\Delta$ and constants $\nu, A_n(\delta)$ of Definition~\ref{def:mbg} can be chosen independently of $t$.
\item[(iv)] that $0$ is almost surely not an eigenvalue of $h_\omega(t)$ tor each $t\in T$.
\end{enumerate}
Then the family $t\in T \mapsto H(t)=\int_{\Omega}^\oplus H_\omega(t)$ with $H_\omega(t)=\sgn(h_\omega(t))$ is continuous on average in the sense of Definition~\ref{def:contonaverage}. 
\end{lemma}
\noindent{\bf Proof.}
The existence of the $L^1(\Omega)$-functions $f(t)$ and the uniform upper bound on their $L^1$-norm is shown in Lemma~\ref{lemma:comm_est_tech}. 

The operator norm $\lVert h_\omega(t)\rVert$ has a uniform upper bound due to (i) since it can be estimated in terms of those matrix elements and the direct integral $h(t)=\int_{\Omega}^\oplus h_\omega(t) \mathrm{d}\PM(\omega)$ obtained from the measurable family is therefore a bounded operator for each $t$. 

It remains to show that $H(t)$ is continuous in the weak operator topology. For that, it is sufficient to show that $h(t)$ depends on $t$ continuously in the strong operator topology. The absence of eigenvalue at $0$ demanded in (iv) then implies continuity in the strong operator topology of spectral projections onto spectral intervals ending at $0$ \cite[VIII.24]{ReedSimon}, which is stronger than weak operator continuity.

The vectors $\delta_y \otimes e \otimes \psi$ for $y\in \Ll$, $e\in \CM^n$ and $\psi\in L^2(\Omega) \cap L^\infty(\Omega)$ span a total subset of $\widehat{\Hh}:=\ell^2(\Ll)\otimes \CM^n \otimes L^2(\Omega)$. Due to the uniform bound on the norms it is enough to prove that each map of the form $t \in T\mapsto h(t)(\delta_y\otimes e \otimes \psi)$ is continuous. 

Since $h(t)$ is a direct integral one has $h(t) (\delta_y \otimes e  \otimes \psi) = M_\psi h(t) (\delta_y  \otimes e \otimes \one_\Omega)$ with the constant $1$-function on $\Omega$ and the multiplication operator $M_\psi$ on $L^2(\Omega)$. This implies
$$\norm{h(t)(\delta_y  \otimes e\otimes \psi)- h(t')(\delta_y  \otimes e\otimes \psi)}_{\widehat{\Hh}} \leq \norm{(h(t)-h(t'))(\delta_y \otimes e\otimes \one_\Omega)}_{\widehat{\Hh}} \, \norm{\psi}_{L^\infty(\Omega)}.$$
Next we insert the partition of unity $\one = \sum_{x\in \Ll}|x\rangle\langle x|$ and apply the triangle inequality:
\begin{align*}\norm{(h(t)-h(t'))(\delta_y \otimes e \otimes \one_\Omega)}_{\widehat{\Hh}}&\leq \sum_{x\in \Ll} \norm{|x\rangle \langle x| (h(t)-h(t'))(\delta_y \otimes e\otimes  \one_\Omega)}_{\widehat{\Hh}}\\
&\leq \sum_{x\in \Ll}\left( \int_{\Omega}  \norm{\langle x| \left( h_\omega(t)-h_\omega(t')\right)|y\rangle}_{M_n(\CM)}^2 \, \norm{e}^2\difd\PM(\omega)\right)^{\frac{1}{2}}
\end{align*}
For the second inequality one writes out the norm of $\widehat{\Hh}$ in terms of the scalar product and then trivially estimates the inner scalar product of $\ell^2(\Ll)\otimes \CM^n$. That each term of the sum converges to $0$ individually as $t\to t'$ is precisely the assumption of $L^2(\Omega)$-continuity of (ii). Due to (i) we can apply dominated convergence to conclude that the sum as a whole converges to $0$. 

\hfill $\Box$

This completes the proof of Theorem~\ref{th:main} since Corollary~\ref{cor:rescaled_dirac_commutator} and Lemma~\ref{lemma:contmbgfamily} together show that the conditions of Proposition~\ref{prop:continuity} are satisfied for any rescaled Dirac operator. Therefore the spectral localizer formalism gives us continuity of the probability distribution of indices. 

\section{Translation-invariance and ergodicity}
\label{sec:tr_inv}

In the definition of $\DD$ we originally shifted the position operators by an arbitrary offset to make sure that $\DD$ is invertible. We can more systematically track this dependence by defining for any $x\in \bbR^d\setminus \Ll$ the shifted Dirac operator
$$\DD_x=\sum_{j=1}^d \gamma_j \otimes (X_j-x_j)$$
centered in $x$.  It is easy to show that the index pairings do not depend on this shift:
\begin{proposition}
	\label{prop:indextranslationinv}
Let $H$ be as in Lemma~\ref{lemma:fermiprojectiondecay} and assume that each $H_\omega$ is of symmetry type $\sigma$. We then consider the (almost surely) Fredholm operators $F_{x,\omega}$ associated to the index pairing between $\DD_x$ and $H_\omega$ in symmetry class $\symlabel$. Then almost surely $\Ind_\symlabel(F_{x,\omega})$ does not depend on $x\in \bbR^d \setminus \Ll$.
\end{proposition}
\noindent{\bf Proof.} 
As each $\DD_x$ has a compact resolvent and $\DD_x-\DD_y$ is a bounded operator basic perturbation theory implies that $\sgn(\DD_x)-\sgn(\DD_y)$ is a compact operator for all $x,y\in \RM^d\setminus \Ll$ (this is also easy to see directly since it is a multiplication operator). Then $F_{x,\omega}-F_{y,\omega}$ is also compact in any symmetry class. Therefore the index pairings do not depend on $x$ (note that for the $\ZM_2$-index invariance under compact perturbations  is proven in \cite{SB15}).
\hfill $\Box$

If the index almost surely does not depend on the configuration $\omega$ then the continuity result of Theorem~\ref{th:main} improves to almost sure constancy of the index. A well-known sufficient condition (dating back at least to \cite{BES}) is the existence of an ergodic translation action:

\begin{proposition}
\label{prop:ergodic}
Let $H$ be as in Proposition~\ref{prop:indextranslationinv}. We say that $H$ is ergodic if there exists a measurable ergodic group action $T: G\times \Omega \to \Omega$ under which $H_\omega$ is covariant in the sense that
$$H_{T_g\omega} = U_g H_\omega U_g^*$$
for a family of unitary operators $(U_g)_{g\in G}$ with the transformation behaviour $U_g^* \DD U_g = \DD_{f(g)}$ for some function $f:G\to \bbR^d\setminus \Ll$. 

If $H$ is ergodic then $\Ind_\symlabel(F_\omega)$ is almost surely constant.
\end{proposition}
\noindent{\bf Proof.}
Ergodicity means that any measurable function that is $G$-invariant up to sets of measure zero is almost surely constant. It is therefore enough to show that almost surely $\Ind_\symlabel(F_\omega)=\Ind_\symlabel(F_{T_g\omega})$ holds for all $g\in G$. For any of the operators \eqref{eq:freds1} or \eqref{eq:freds2} one has
$$F_{T_g\omega} = U_g F_{f(g),\omega} U_g^*$$
such that invariance of the index under unitary transformations means that it is enough to show that $F_\omega$ has the same index as $F_{f(g),\omega}$ and we already saw that in Proposition~\ref{prop:indextranslationinv}. $\Box$

The prototypical example are the (magnetic) translations on $\ell^2(\bbZ^d)$ which are a (projective) representation of $\ZM^d$. For example, the Hamiltonian \eqref{eq:example_ham_disorder} is ergodic in this sense. 

From Corollary~\ref{cor:almost_surely_constant} we can then immediately conclude:
\begin{corollary}
In the setting of Theorem~\ref{th:main} assume that each $h(t)$ is ergodic in the sense of Proposition~\ref{prop:ergodic}. Then the index $\Ind_\sigma(F_\omega(t))$ almost surely does not depend on $\omega$ and the almost sure value is locally constant with respect to $t$.
\end{corollary}
This provides a large class of models for which we know that the strong topological invariant is constant on any path that preserves the mobility gap condition.

\section{Delocalization of interfaces}
\label{sec:interfaces}

Consider two (random families of) Hamiltonians $h^{(1)}$ and $h^{(2)}$ on a pattern $\Ll$, both symmetric under the same symmetry operators and with a mobility gap in an common interval $\Delta$. If one partitions $\Ll = \Lambda_1 \cup \Lambda_2$ into two suitably shaped infinite regions one can construct an interface Hamiltonian $h^{(I)}$ which looks like $h^{(1)}$ far away from $\Lambda_2$ and like $h^{(2)}$ far away from $\Lambda_1$, for example if $\Lambda_1 = \ZM^{d-1}\times \NM$ and $\Lambda_2 = \ZM^{d-1}\times (-\NM)$. When the two asymptotic {\it bulk} Hamiltonians have different strong topological invariants then $h^{(I)}$ should have interface states in the spectral interval $\Delta$ which are localized to the interface region and which carry a topological charge corresponding to the difference of the bulk invariants. 

Indeed, this can be proven in the framework of K-theoretic bulk-edge correspondence, which goes back to \cite{KellendonkRMP2002} and has since then been generalized to all dimensions and symmetry classes, see in particular the references \cite{Kubota, BKR, AMZ, PSbook,EwertMeyer} for a far from exhaustive list. However, those results mostly only apply if the bulk Hamiltonians have an actual spectral gap, not just a mobility gap. In the latter case much less is known, though there are some bulk-boundary correspondence results in low dimensions \cite{EGS, ShapiroGraf2018, BSS, SSt, Thesis} which do not use K-theory and also apply in the mobility-gapped setting. It is commonly believed that such interface states are localized to the interface region but must propagate along it in some parallel directions. If the interface region is infinitely extended this presumably coincides with the occurrence of absolutely continuous spectrum inside the bulk gap. Proving the latter for all dimensions and symmetry classes seems out of reach at this point, however, what is more accessible is to show via contradiction that the fractional moments bound of Definition~\ref{def:mbg} cannot hold. The simple idea is that if the interface Hamiltonian $h^{(I)}$ itself has a mobility gap then it also has a well-defined index. Since that index is the same for the Dirac operator $\DD_x$ centered at any point $x\in \RM^d\setminus\Ll$ and is locally computable using the spectral localizer, it must simultaneously coincide with the index of $h^{(1)}$ and the index of $h^{(2)}$. If one glues two bulk Hamiltonians with different index then the resultant Hamiltonian therefore cannot have a mobility gap. 
This argument works the same for more than two asymptotic Hamiltonians, for generality, it is therefore convenient to consider an interface Hamiltonian $h^{(I)}$ which looks like some {\it bulk} Hamiltonian $h^{(B)}$ far away from some {\it interface region} $\Lambda\subset \Ll$.

A technical difficulty, however, is that while the spectral localizer is extremely local in space, it is expressed in terms of the band-flattened $H^{(*)}=\sgn(h^{(*)})$ instead of $h^{(*)}$, $*\in \{I,B\}$, and going over to the spectral flattening could introduce non-locality. Therefore, we need to show that this does not happen as long as the interface Hamiltonian is also mobility-gapped.

\begin{lemma}
\label{lemma:mbgperttech}
Let $h^{(I)}$ and $h^{(B)}$ be random families of Hamiltonians on $\ell^2(\Ll)\otimes \CM^n$ that are short-range, have a mobility gap around $0$ and for which $0$ is almost surely not an eigenvalue, in the same sense as (i), (iii), (iv) of Lemma~\ref{lemma:contmbgfamily}.

Let $\Lambda \subset \Ll$ be some set and assume that $V=h^{(B)}- h^{(I)}$ is short-range and local to $\Lambda$ the sense that for each $k\in \bbN$ there is a constant $C_k$ such that
$$\sup_{\omega\in\Omega}\norm{\langle x| V_\omega | y \rangle}_{M_n(\bbC)} \leq  C_k \langle \mathrm{dist}(x, \Lambda)\rangle^{-k} \langle\mathrm{dist}(y, \Lambda)\rangle^{-k} \langle x-y\rangle^{-k}$$
holds uniformly for all $x,y\in \Ll$.

If both $h^{(I)}$ and $h^{(B)}$ have a mobility gap as in Definition~\ref{def:mbg} then we have a bound
$$\int_{\Omega} \norm{\langle x| H^{(I)}_\omega - H^{(B)}_\omega| y\rangle}_{M_n(\bbC)} \mathbb{P}(\difd{\omega}) \leq \tilde{C}_k\langle \mathrm{dist}(x, \Lambda)\rangle^{-k} \langle\mathrm{dist}(y, \Lambda)\rangle^{-k}$$
for each $k\in \bbN$ and some constant $\tilde{C}_k$.
\end{lemma}
\noindent{\bf Proof.}
We may assume that $h^{(I)}$ and $h^{(B)}$ satisfy the mobility gap condition for the same constants, in particular the same fractional exponents $0 < \nu < 1$ since one can always make $\nu$ smaller while keeping the faster-than-any-polynomial decay. 
Next, we note that in \eqref{eq:fractional_moments} one can replace the operator norm by the $\nu$-Schatten quasi-norm on $M_n(\CM)$; indeed, if the inequality holds for the operator norm then it also holds for the quasi-norm as they are equivalent for finite-dimensional matrices. 
Thus we can assume a fractional moments bound of the form $$\norm{\langle x|\frac{1}{h^{(*)} -\eta}|y\rangle}^{\nu}_{L_n^{\nu}(\Omega)}\leq A_{\nu, k}(\delta) \langle x-y\rangle^{-k} $$
for any $k\in \NM$ with $L_n^{\lambda}(\Omega)$ the $L^{\lambda}(\Omega)$-averaged Schatten norm
$$\norm{\int_\Omega^\oplus a_\omega \difd \PM(\omega)}^\lambda_{L^\lambda_n(\Omega)} = \int_{\Omega} \mathrm{tr}(\abs{a_\omega}^\lambda)\difd \PM(\omega).$$ It is well-known that $L_n^{p}(\Omega)=M_n(\CM)\otimes L^p(\Omega)$ is a $p$-Banach space for $0<p<1$, indeed, it is the non-commutative $L^p$-space associated to the von Neumann algebra $M_n(\CM)\otimes L^\infty(\Omega)$. In particular, the quasi-norms also satisfy the usual H\"older inequality (see \cite{Pisier03} for basic facts on those spaces).

Comparing the resolvents one has
$$\frac{1}{h^{(I)}_\omega - \eta} - \frac{1}{h_\omega^{(B)} - \eta} = \frac{1}{h^{(I)}_\omega -\eta} V_\omega \frac{1}{h_\omega^{(B)} - \eta} $$
such that one can expand the matrix elements as follows
$$\langle x|\left(\frac{1}{h^{(I)}_\omega -\eta}-\frac{1}{h^{(B)}_\omega -\eta}\right)|y\rangle= \sum_{z_1,z_2\in \Ll} \langle x| \frac{1}{h^{(I)}_\omega -\eta}|z_1\rangle\, \langle z_1| V_\omega |z_2\rangle\, \langle z_2| \frac{1}{h^{(B)}_\omega -\eta}|y\rangle.$$
The Hölder inequality for non-commutative $L^p$-spaces implies
$$\norm{f_1f_2f_3}_{L_n^{\frac{\nu}{2}}(\Omega)}\leq \norm{f_1}_{L_n^{\nu}(\Omega)}\,\norm{f_2}_{L_n^{\infty}(\Omega)}\norm{f_3}_{L_n^{\nu}(\Omega)}$$
such that the fractional moments satisfy for any $k\in \NM$ an estimate
\begin{align*}
&\norm{\langle x|\left(\frac{1}{h^{(I)} -\eta}-\frac{1}{h^{(B)} -\eta}\right)|y\rangle}^{\frac{\nu}{2}}_{L_n^{\frac{\nu}{2}}(\Omega)}\\ &\leq c_{k} \sum_{z_1,z_2\in \Ll} \langle x-z_1\rangle^{-k} \langle z_1-z_2\rangle^{-k} \langle \mathrm{dist}(z_1, \Lambda)\rangle^{-k} \langle\mathrm{dist}(z_2, \Lambda)\rangle^{-k} \langle z_2-y\rangle^{-k},
\end{align*}	
It is then elementary to conclude that one also has for any $k\in \NM$ an estimate
$$\int_{\Omega} \norm{\langle x| \left(\frac{1}{h^{(I)} -\eta}-\frac{1}{h^{(B)} -\eta}\right)| y\rangle}^{\frac{\nu}{2}}_{M_n(\bbC)} \mathbb{P}(\difd{\omega}) \leq \tilde{c}_k \langle \mathrm{dist}(x, \Lambda)\rangle^{-k} \langle\mathrm{dist}(y, \Lambda)\rangle^{-k}$$
that holds uniformly for all $\eta$ which satisfy $\mathrm{dist}(\eta, \mathrm{Spec}(h^{(*)})\setminus \Delta)>\delta$ for some small enough $\delta>0$ as prescribed in Definition~\ref{def:mbg}. The same contour integration argument as in Lemma~\ref{lemma:fermiprojectiondecay} transfers this bound to the averaged difference of $H_\omega^{(I)}$ and $H_\omega^{(B)}$.
\hfill $\Box$

We will now want $\Ll\setminus \Lambda$ to contain balls of arbitrary size which are arbitrarily far away from $\Lambda$ such that one can properly take a bulk limit. For technical reasons we need to impose a more precise growth condition: 

\begin{definition}
We say that $\Lambda \subset \Ll$ allows a bulk limit if $\Ll \setminus \Lambda$ contains a sequence of balls $B_{R_n}(x_n)\cap \Ll$ with $\abs{x_n}\to \infty$, $\mathrm{dist}(x_n, \Lambda) > 2R_n$ and such that $R_n > \abs{x_n}^\xi$ for some fractional power $0<\xi<1$.
\end{definition}
The prototypical example of an interface region is $\Ll=\ZM^d$ and $\Lambda=\ZM^{d-1} \times \NM$ as one can see by choosing balls with radius $\abs{x_n}^{\frac{1}{2}}$ for a sequence $x_n\to \infty$. The reason behind this scaling condition is that to compute the strong topological invariant with the spectral localizer constructed from $\DD^{(r)}_{x_n}$ with probability greater than $1-\epsilon$ one will need to use the matrix elements in a ball centered in $x_n$ whose minimal size increases with $\epsilon^{-1}$ but also mildly with the distance $\abs{x_n}$ of the ball from the origin. The growth condition will ensure that there is for any $\epsilon>0$ some ball in $\Ll \setminus \Lambda$ which is sufficiently large to apply the spectral localizer method and also far enough away from $\Lambda$ such that replacing $H_\omega^{(I)}$ with $H_\omega^{(B)}$ rarely changes the signature.

\begin{proposition}
\label{prop:bulkindex}
Assume the situation of Lemma~\ref{lemma:mbgperttech} and that $\Lambda$ allows a bulk limit, then one has almost surely $$\Ind_\symlabel(F^{(I)}_\omega) = \Ind_\symlabel(F^{(B)}_\omega).$$
\end{proposition}
\noindent{\bf Proof.}
Denote $\DD_{x_n}$ the Dirac operator with center shifted by $x_n$ and its associated weight by $w_{x_n}=(1+\DD_{x_n}^2)^{-\frac{1}{2}}$. Since $H^{(*)}$ commutes with the Clifford degrees of freedom this shift drops in the commutator to yield $$\lVert [\DD_{x_n}, H_\omega^{(*)}] w^s_{x_n}\rVert = \lVert [\DD,H_\omega^{(*)}] w^s w^{-s} w^s_{x_n}\rVert \leq c \lVert [\DD,H_\omega^{(*)}] w^s\rVert\, \langle x_n \rangle^s$$
for both $*\in \{B,I\}$. 

We will want to fix the otherwise arbitrary exponents $0<s<r$ in such a way that we are able to use effectively the spectral localizer $L^\symlabel_{\kappa,\rho}(\DD^{(r)}_x, H_\omega^{(*)})$ based on the shifted rescaled Dirac operator $\DD_x^{(r)}$. By Lemma~\ref{lemma:comm_est_tech} and Corollary~\ref{cor:rescaled_dirac_commutator} we have for any $s>0$ an estimate of the form
$$\lVert [\DD^{(r)}_{x_n}, H_\omega^{(*)}]\rVert\leq f_s(\omega) \langle x_n \rangle^s$$
with an $L^1(\Omega)$-function $f_s$ that does not depend on $n$. By the Markov inequality we can therefore choose two constants $C_1$, $C_2$ depending only on the $L^1$-norm of $f_s$ such that for each $n$ the pair
$$\kappa_n = \epsilon C_1  \langle x_n\rangle^{-s}, \quad \rho_n = \epsilon^{-1}C_2  \langle x_n\rangle^s$$
is admissible as in Theorem~\ref{theorem:skewlocalizer} for the spectral localizer $L^\symlabel_{\kappa_n,\rho_n}(\DD^{(r)}_{x_n}, H_\omega^{(*)})$ with probability greater than $1-\epsilon$, i.e.
$$\bbP\left(\Ind(F_\omega^{(*)}) =\mathrm{Sig}_\symlabel(L^{\symlabel}_{\kappa_n,\rho_n}(\DD^{(r)}_{x_n}, H_\omega^{(*)})\right) > 1-\epsilon,$$
which also uses that the true index does not depend on $x_n$ by Proposition~\ref{prop:indextranslationinv}.

Taking into account the rescaling $L^\symlabel_{\kappa,\rho}(\DD^{(r)}_{x_n}, H_\omega^{(*)})$ is built from matrix elements of $H_\omega^{(*)}$ in a ball of radius less than $\rho_n^{1/(1-r)}$ around $x_n$. Since $\rho_n = O(\langle x_n\rangle^{s})$ and $r,s$ can be made arbitrarily small, we assume that $s/(1-r)< \xi$, which implies that there is for any $\epsilon>0$ some $n_0$ large enough such that $$R_n > \langle x_n\rangle^{\xi} >  \rho_n^{1/(1-r)}$$ for all $n>n_0$ and hence those balls are then contained entirely in $\Ll\setminus \Lambda$ with $\mathrm{dist}(x_n, \Lambda)> 2R_n$. 

Since the number of matrix elements involved in the localizer is bounded polynomially in $\rho_n$ the bound of Lemma~\ref{lemma:mbgperttech} on the average difference of matrix elements implies that there is for each $k\in \NM$ a constant $c_k$ such that $$\mathbb{E}\norm{L^{\symlabel}_{\kappa_n,\rho_n}(\DD^{(r)}_{x_n}, H^{(I)})-L^{\symlabel}_{\kappa_n,\rho_n}(\DD^{(r)}_{x_n}, H^{(B)})}_{M_m(\CM)}\leq \mathbb{E}\norm{P_n(H^{(I)}-H^{(B)})P_n}\leq c_k \langle x_n\rangle^{-k}$$
with $m$ the matrix size of the skew localizer $L^\sigma_{\kappa,\rho}$, $P_n$ the projection to the ball of radius $\rho_n^{1/(1-r)}$ around $x_n$ and where $c_k$ depends on $\epsilon$ but not on $n$.

By Markov's inequality this implies
$$\PM(\norm{L^{\symlabel}_{\kappa_n,\rho_n}(\DD^{(r)}_{x_n}, H^{(I)}_\omega)-L^{\symlabel}_{\kappa_n,\rho_n}(\DD^{(r)}_{x_n}, H^{(B)}_\omega)} \geq \frac{ c_k \langle x_n\rangle^{-k}}{\epsilon})\leq \epsilon.$$
Choose now any $n$ large enough such that $\frac{ c_k \langle x_n\rangle^{-k}}{\epsilon}< \frac{1}{4}$. As argued in the proof of Proposition~\ref{prop:continuity}, if for admissible $\kappa,\rho$ two spectral localizers for Hamiltonians of spectral gap $1$ have norm difference smaller than $\frac{1}{4}$ then they are homotopic and therefore have the same signature. In conclusion, we have for any $\epsilon>0$ some $n$ such that
$$\bbP\left(\Ind_{\symlabel}(F^{(I)}_\omega)=\mathrm{Sig}_\symlabel(L^{\symlabel}_{\kappa,\rho}(\DD^{(r)}_{x_n}, H^{(I)}_\omega)) = \mathrm{Sig}_\symlabel (L^{\symlabel}_{\kappa,\rho}(\DD^{(r)}_{x_n}, H^{(B)}_\omega))=\Ind_{\symlabel}(F^{(B)}_\omega)\right) > 1-2\epsilon.$$

\hfill $\Box$

For any two bulk Hamiltonians $h^{(1)}, h^{(2)}$ one can construct an interface Hamiltonian $h^{(I)}$ by restricting them to the subspaces $\ell^2(\Lambda)\otimes \CM^n$ and $\ell^2(\Ll\setminus \Lambda)\otimes \CM^n$ respectively via projections $P_\pm$ and setting 
$$h^{(I)}_\omega =P_+ h^{(1)}_\omega P_+ + P_- h^{(2)}_\omega P_- + \hat{V}_\omega$$ 
where the coupling term $\hat{V}$ is mostly arbitrary as long as it decays with the distance to the boundary of $\Lambda$. If both $\Lambda$ and $\Ll\setminus \Lambda$ are large enough to admit a bulk limit and the two bulk Hamiltonians have different index with positive probability then we can conclude that the resulting interface Hamiltonian will not have a mobility gap (except possibly when $0$ is an eigenvalue, see below).

The result gives new information even in the rather well-understood case where the bulk Hamiltonians have an actual spectral gap. Then the machinery of $K$-theory implies that an interface of two Hamiltonians in the same symmetry class but different bulk topological invariant must have in-gap states which are protected by an edge topological invariant (see \cite{PSbook} and for the real AZ-classes \cite{AMZ, Kubota, BKR, EwertMeyer}). Here the subtle condition of Lemma~\ref{lemma:mbgperttech} that $h^{(I)}$ does not have $0$ as an eigenvalue becomes relevant: In one dimension the interface states are finitely degenerate isolated zero-modes inside the bulk gap protected by a chiral or particle-hole symmetry. Those in-gap states are always localized since there are only finitely many of them. This is no contradiction, since the interface Hamiltonian does not even have a well-defined index if it has $0$ as an eigenvalue and thus the argument above does not apply. In dimension two or higher the interface states are generically carried by a larger spectral interval and there is no reason for $0$ to be an almost sure eigenvalue, therefore this is no longer a significant limitation. In dimension two in the complex symmetry classes $A$ and the odd time-reversal protected class $AII$ one can in the spectrally gapped case further show that, at least in a half-space geometry, the interface states imply the existence of absolutely continuous spectrum inside the bulk gap \cite{BolsWerner,BolsCedzich}. Those results are based on a classification of certain unitary operators up to trace-class perturbation and does not easily translate to boundary states of different dimension. While the results here are not as fine as to establish absolutely continuous spectrum, the impossibility of the fractional moments bound to hold suggests absence of dynamical localization under any reasonable definition.

\appendix
\section{Clifford algebras and their symmetries}
\label{sec:app}
The complex Clifford algebra $\CM_d$ of $d$ generators is the universal complex algebra generated by $d$ self-adjoint elements $\gamma_1,...,\gamma_d$ such that $\{\gamma_i,\gamma_j\}=2 \delta_{i,j}$. In this paper we use the following irreducible representation of $\CM_d$ on $M_{2^{\lfloor d/2\rfloor}}(\CM)$ constructed using recursion with the same conventions as in \cite{PSbook}. Given the representation $(\sigma_1,...,\sigma_{d})$ of the generators of $\CM_{d}$ with $d$ odd then one obtains a representation of $\CM_{d+2}$ by setting 
$$
\gamma_i 
\;=\; 
\begin{pmatrix}
	0 & \sigma_i \\ \sigma_i & 0
\end{pmatrix}
\quad 
\forall\; \,
i=1,\ldots,d\,, 
\qquad 
\gamma_{d-1} 
\;=\; 
\begin{pmatrix}
	0 & \imath \\ -\imath & 0
\end{pmatrix}
\;,
\quad 
\gamma_{d} 
\;=\; 
\begin{pmatrix}
	\one & 0 \\ 0 & -\one
\end{pmatrix}
\;.
$$
The starting point of the recursion is $\sigma_1=1$ for $d=1$. For any even $d$ one obtains an irreducible representation $\CM_{d}$ by omitting the last generator of $\CM_{d+1}$. 

For any representation like this, where the even generators $\gamma_{2k}$ are imaginary and the odd generators $\gamma_{2k+1}$ are real with respect to the usual complex conjugation, the real symmetries are well-understood \cite{GS16}: One can define the following real self-adjoint unitaries
$$\Omega=\imath^{\lfloor d/2\rfloor} \gamma_1 ...\gamma_d, \quad \symlabel = \imath^{\lfloor d/2\rfloor} \gamma_2 \gamma_4 ... \gamma_{2\lfloor d/2\rfloor}, \quad \widehat{\Sigma} =  (-1)^{\lfloor d/2\rfloor}\gamma_1 \gamma_3 ... \gamma_{2\lfloor d/2\rfloor-1}.$$
A Clifford vector $X=x\cdot \gamma$ with real entries $x\in \RM^d$ will satisfy the symmetry relation
$$\Sigma^* \overline{X} \Sigma = (-1)^{\lfloor d/2\rfloor}  {X}.$$ 
For even $d$ one has in addition the symmetry relations
$$\widehat{\Sigma}^* \overline{X} \widehat{\Sigma} = -(-1)^{\lfloor d/2\rfloor}  {X}$$
and
$$\Omega^* X \Omega = -X.$$ 
In the chosen representation the self-adjoint unitary $\Omega$ is proportional to the chiral symmetry operator $\gamma_{d+1}=\mathrm{diag}(\one,-\one)$ and the other symmetry operators also have expressions in terms of tensor products of Pauli matrices. In even $d$ one has two real symmetry operators $\Sigma$, $\widehat{\Sigma}$ which either commute or anti-commute depending on $d$. Together with the complex conjugation and a factor of $\imath$ if necessary one can therefore define from them two commuting anti-unitary operators which encode the real symmetries of the Clifford vectors.

\section{Regularization of Dirac operators}
\label{sec:app2}

As was noted originally in \cite{Kaad2020} one can improve the regularity of a spectral triple by using instead of $\DD$ the rescaled operator $\DD^{(r)} = \DD (1+\DD^2)^{-\frac{r}{2}}$ for some fractional power $0<r<1$. We will prove in this section that the commutator $[\DD^{(r)},H]$ is bounded if $[\DD,H]$ is relatively bounded w.r.t. a fractional power of $\DD$ smaller than $\DD^{(r)}$. 

\begin{proposition}
\label{prop:regularisation}
 Let $\DD$ be a self-adjoint operator densely defined on a Hilbert space $\Hh$ and let $H$ be a bounded self-adjoint operator on $\Hh$ which maps the core $\Ee = \cap_{k\in \NM} \mathrm{Dom}(\DD^k)$ of $\DD$ into itself $H\Ee\subset \Ee$. With the bounded weight $w:=(1+\DD^2)^{-\frac{1}{2}}$ assume that there is a constant $0<s<1$ such that $$\|[\DD,H]w^{s}\| < A_s$$ in the sense that this operator extends from $\Ee$ to a bounded operator, then the commutator $$[\DD^{(r)}, H]$$ with $\DD^{(r)}= \DD w^r =\DD (1+\DD^2)^{-\frac{r}{2}}$ extends from $\Ee$ to a bounded operator with
	$$\norm{[\DD^{(r)}, H]} \leq c_{r,s} A_s$$
 for all $s<r<1$ and some universal constant $c_{r,s}$ independent of $H$ and $\DD$.
\end{proposition}

We will use a similar argument as in the proof of  \cite[Lemma 25]{SSt2} which uses the Helffer-Sj\"ostrand calculus for smooth functions \cite{Davies95}. For $\rho \in \RM$, let $\Ss^{\rho}(\RM)$ denote the set of smooth functions   $f:\RM\to\RM$ satisfying
$$
\lvert\partial^k f(x)\rvert \;\leq\; C_k \langle x\rangle^{\rho-k}
\;, \qquad k\in \NM\;
$$
for $\langle x\rangle =(1+x^2)^{\frac{1}{2}}$. For any $N>0$ there is an almost analytic extension $\tilde{f}_N$ of $f$ to $\CM$  which satisfies
\begin{equation}
	\label{eq:hsbounds}
	\lvert \partial_{\overline{z}} \tilde{f}(x+\imath y) \rvert 
	\;\leq\; c_N \left(\sum_{k=1}^{N+1} C_k\right)  \langle x\rangle^{{\rho-1-N}}\, \lvert y\rvert^{N}
\end{equation}
for a universal constant $c_N$ and is supported in the region $G=\{x+\imath y: \,\lvert y\rvert < 2\langle x\rangle\}$. For any $\rho < 0$ and $N\geq 1$, the smooth functional calculus can be written as a norm-convergent integral
\begin{equation}
	\label{eq:hsrep}
	f(\DD) \;=\; \frac{1}{\pi}\int_G (\partial_{\overline{z}} \tilde{f}_N(z)) (\DD-z)^{-1} \mathrm{d}z \wedge \mathrm{d}\overline{z}.
\end{equation}

\noindent{\bf Proof (of Proposition~\ref{prop:regularisation})}

\vspace{.1cm}
Set $f(x)=x (1+x^2)^{-\frac{r}{2}}$ and note that $f \in \Ss^\rho(\RM)$ for $\rho = 1-r$.
Choose any smooth function $\chi$ equal to $1$ on $[-1,1]$ and vanishing outside $(-2,2)$ and regularize $f_R(\lambda)=f(\lambda) \chi(\lambda R^{-1})$. There is for any $k\in \NM$ a constant $c_k$ such that $\abs{\partial^k \chi(\lambda)} \leq c_k \langle \lambda\rangle^{-k}$ and then by scaling 
$$
\abs{\partial^k f_{R}} 
\;\leq\; 
\sum_{m=0}^k C_m \langle x\rangle^{\rho-k}
\,\frac{1}{R^{k-m}}\, c_{k-m} 
\;\leq\; \widehat{C}_k 
\langle x\rangle^{\rho-k}
$$
with constants uniformly in $R \geq 1$. With an almost analytic extension $\tilde{f}_{R,N}$ and $\psi \in \Ee$, one can then write
\begin{align*}
	[f_R(\DD),H]\psi
	&
	\;=\; \frac{1}{\pi}\int_G (\partial_{\overline{z}} \tilde{f}_{R,N}(z)) \,[(\DD-z)^{-1},H]\psi \,\mathrm{d}z \wedge \mathrm{d}\overline{z}
	\\
	&
	\;=\; 
	-\frac{1}{\pi}\int_G (\partial_{\overline{z}} \tilde{f}_{R,N}(z))\, (\DD-z)^{-1}[\DD,H](\DD-z)^{-1}\psi\, \mathrm{d}z \wedge \mathrm{d}\overline{z}	\\
	&
	\;=\; 
	-\frac{1}{\pi}\int_G (\partial_{\overline{z}} \tilde{f}_{R,N}(z))\, (\DD-z)^{-1} [\DD,H]w^{s}\,w^{-s}(\DD-z)^{-1}\psi\, \mathrm{d}z \wedge \mathrm{d}\overline{z}.
\end{align*}
For $z=x+\imath y$ one may estimate
$$\norm{(\DD-z)^{-1}}\leq \abs{y}^{-1}$$
and by functional calculus and some elementary analysis
\begin{align*}
\norm{w^{-s}(\DD-z)^{-1} } &= \sup_{d\in \RM}\abs{(1+d^2)^{\frac{s}{2}}(d-z)^{-1} } \\
&=\sup_{d\in \RM}{(1+d^2)^{\frac{s}{2}}((d-x)^2+y^2)^{-\frac{1}{2}}}\\
&\leq\sup_{d\in \RM}{(1+d^2)}^{\frac{s}{2}}((d-x)^2+y^2)^{-\frac{s}{2}}  \abs{y}^{-1+s}\\
&=\left(\frac{1+x^2+y^2 +\sqrt{4x^2+ (x^2+y^2-1)^2}}{y^2}\right)^{\frac{s}{2}} \,\abs{y}^{-1+s}\\
&\leq c\left(1+\frac{1+x^2}{y^2}\right)^{\frac{s}{2}} \,\abs{y}^{-1+s}\leq  c\,\abs{y}^{-1+s} +c \langle x\rangle^{s} \,\abs{y}^{-1}.
\end{align*}
From \eqref{eq:hsbounds} we conclude that
$$
\norm{(\partial_{\overline{z}} \tilde{f}_{R,N}(z)) (\DD-z)^{-1}[\DD,H](\DD-z)^{-1}}
\;\leq\; 
\tilde{c}_N(\langle x\rangle^{\rho-1-N}\abs{y}^{N-2+s}+ \,\langle x\rangle^{\rho+s-1-N} \,\abs{y}^{N-2})
$$
with constants that are uniform on $G$ and do not depend on $R$. If the right hand side has a finite integral over $G$ then the proof is complete, since convergence of $f_R(\DD)\psi$ to $f(\DD)\psi$ for each $\psi\in \Ee$ implies then that the norm of the commutator $[f(\DD),H]$ is bounded by that same uniform upper bound. Assuming $N\geq 2$ one can first integrate over $y$ to get an expression proportional to $\langle x\rangle^{\rho+s-2}$ and therefore the integral over $x$ exists if $\rho+s-2 < -1$ which is equivalent to $s<r$.
\hfill $\Box$

\vspace{.2cm}
\noindent\small{{\bf Acknowledgements}\;\; The author thanks Jacob Shapiro, Emil Prodan, Nora Doll and Hermann Schulz-Baldes for helpful discussions. This research was supported by the German Research Foundation (DFG) Project-ID 521291358. }
\vspace{.2cm}

\noindent\small{{\bf Conflict of interest}\;\; The author has no conflict of interest to declare.}
\vspace{.2cm}

\noindent\small{{\bf Data Availability Statement}
Data sharing is not applicable since no data was created or analyzed.}

%%%%%%%%%%%%%%%%%%%%%%%%%%%%%%%%%%%%%%%%%%%%%

\end{document}